\documentstyle[amssymb,aps,prb,epsf]{revtex}
\begin{document}
\author{A.\ E.\ Koshelev and I.\ Aranson}
\address{Materials Science Division, Argonne National Laboratory,
Argonne, Illinois 60439}
\title{Dynamic structure selection and instabilities of driven
Josephson lattice in high-temperature superconductors}
\date{\today}
\maketitle

\begin{abstract}
We investigate the dynamics of the Josephson vortex lattice in layered
high-T$_{c}$ superconductors at high magnetic fields.  Starting
from coupled equations for superconducting phases and
magnetic field we derive equations for the relative displacements
[phase shifts] between the planar Josephson arrays in the
layers.  These equations reveal two families of steady-state
solutions: lattices with constant phase shifts between
neighboring layers, starting from zero for a  rectangular
configuration to $\pi$ for a triangular configuration, and
double-periodic lattices. We find that the excess Josephson current is
resonantly enhanced when the Josephson frequency matches the
frequency of the plasma mode at the wave vector selected by
the lattice structure.  The regular lattices exhibit several kinds of
instabilities.  We find stability regions of the moving lattice
in the plane lattice structure - Josephson frequency.
A specific lattice structure at given velocity is selected uniquely by
boundary conditions, which are determined by the reflection properties of
electromagnetic waves generated by the moving lattice.  With increase
of velocity the moving configuration experiences several qualitative
transformations.  At small velocities the regular lattice is stable and
the phase shift between neighboring layers smoothly decreases with
increase of velocity, starting from $\pi$ for a static lattice.  At the
critical velocity the lattice becomes unstable.
At even higher velocity a regular lattice is restored again with
the phase shift smaller than $\pi/2$.
With increase of velocity, the structure evolves towards a rectangular
configuration.
\end{abstract}


\section{Introduction}

The intrinsic Josephson effect \cite{IntrJos} in strongly anisotropic
high-temperature superconductors, such as
Bi$_{2}$Sr$_{2}$CaCu$_{2}$O$_{x}$ (BSCCO), has been a subject of
intense experimental study (see review articles
\cite{IJJReviews,YurgensReview}). Stack of low-dissipative
junctions formed by atomic layers represents a nonlinear system with
unique dynamic properties.  In particular, this system possesses a
rich spectrum of the electromagnetic plasma waves. These waves can
be excited via the Josephson effect, which makes these materials
convenient voltage-to-frequency converters and gives strong
potential for practical applications.

A magnetic field applied along the layers forms the lattice of
Josephson vortices.  Transport properties of this state are
determined by the dynamics of the Josephson vortex lattice.  Coherent
motion of Josephson vortices in BSCCO has been observed
experimentally \cite{LeeAPL95,HechPRL97,LatyshPhysC97}.  The motion
of the Josephson lattice generates alternating electric fields
and currents which are coupled to electromagnetic plasma waves.
One can expect a strong resonance emission when the velocity of the
lattice matches the velocity of the plasma wave \cite{KoySSC95}.  In
a voltage-biased transport experiment this should be seen as a resonant
enhancement of the transport current \cite{BulPRB96,ArtJETPL97}, while in
a current-biased experiment the resonance should be seen as a voltage
jump.  In conventional long Josephson junctions this effect is
known as Eck peak (Eck step) \cite{Eck} (see also recent article
\onlinecite{CirilloPRB98} and references therein). However, in
layered superconductors this effect has qualitatively new
features.  The velocity of a plasma wave propagating along the
layers strongly depends on the wave vector along the $c$
direction.  On the other hand, the traveling Josephson lattice is
coupled only to the plasma modes with the wave vectors given by
the reciprocal wave vectors of the lattice, i.e., \emph{the wave
vector of the resonant mode is determined by the structure of the
Josephson lattice} \cite{SakaiPRB94}.  In early theoretical work
\cite{KoySSC95,BulPRB96,ArtJETPL97} it was assumed that the moving
lattice preserves its static structure (stretched triangular
lattice) and the resonant velocity of the lattice is determined by
this structure.  We will show that the lattice structure
experiences a nontrivial evolution with increase of velocity.
Moreover, a regular lattice state always becomes unstable at a
certain velocity.  Therefore, finding conditions for the resonance
emission is a nontrivial problem. Recent numerical studies
\cite{Mach,KleinPRB00} revealed a very rich dynamic evolution of
moving Josephson structures, which includes regular as well as
chaotic solutions.

In this paper we analyze the dynamics of the Josephson lattice at strong
fields, in the regime of strongly overlapping vortices. In this
regime Josephson coupling can be treated perturbatively. This
strongly facilitates the analytical analysis of the dynamic
structures. Major results have been already reported in the Letter
\onlinecite{jldPRL00}. A similar approach has been used to calculate
resonant current-voltage dependencies in conventional long
Josephson junctions in magnetic
field.\cite{Eck,KulikJETPLett65,CirilloPRB98} This approach also
describes Josephson dynamics at arbitrary magnetic field when all
junction are driven into resistive state by transport current. We
will focus on the dynamics in the case of large size samples and
will not consider resonances related to finite-size effect (Fiske
resonances). The classification of Fiske resonant frequencies for
layered superconductors in the case of a moving rectangular lattice
has been done by Kleiner. \cite{KleinerPRB94A} The experimental
observation of these modes has been reported by Irie {\emph et.
al.}\cite{IrieAPL98}

The phase dynamics in layered superconductors is described by the
coupled sine-Gordon
equations.\cite{SakaiJAP93,SakaiPRB94,KleinerPRB94,BulPRB94,BulPRB96,ArtJETPL97}
Two equivalent approaches have been used to derive these equations.  In
the first approach \cite{SakaiJAP93,SakaiPRB94,KleinerPRB94} the layered
superconductor is modeled as a multilayered S-I-S-I-\ldots \ system and
the parameters of the equations are expressed via the geometrical parameters and
the bulk parameters of the superconducting material from which the superconducting
layers are prepared.  The second
approach\cite{BulPRB94,BulPRB96,ArtJETPL97}, more natural for atomically
layered superconductors, is based on the Lawrence-Doniach model.  In
this approach the layers are treated as two-dimensional entities from
the very beginning.  We will use the second approach.

The dynamic phase diagram strongly depends on dissipation mechanisms.
Moving Josephson vortices generate both in-plane and out-of-plane
electric fields which induce dissipative quasiparticle currents.
Usually, only dissipation due to the tunneling of quasiparticles
between the layers is taken into account in consideration of
the dynamics of the Josephson vortices.\cite
{ClemCoffPRB90,SakaiJAP93} However, in high-T$_{c}$ superconductors
the in-plane component of the quasiparticle conductivity $\sigma
_{ab}$ is strongly enhanced in the superconducting state as compared
to the normal conductivity due to the reduction of the phase space for
scattering.  This enhancement was observed in $YBa_2Cu_3O_7$
\cite{SigmaQYBCO} and in $BSCCO$\  \cite{SigmaQBSCCO}.  The c-axis
component, $\sigma _{c}$, monotonically decreases with temperature
in the superconducting state.\cite{Sigmac,LatyshevPRL99} The
anisotropy of dissipation $\sigma _{ab}/\sigma _{c}$ becomes
larger than the superconducting anisotropy $\gamma ^{2}$ (here
$\gamma $ is the anisotropy of the London penetration depth). This
leads to the dominance of the in-plane dissipation in the dynamics of
the Josephson lattice in a wide field range.\cite{KoshPRB00}

Starting from the the coupled sine-Gordon equations, we derive equations
for the relative displacements between the moving Josephson planar
arrays in the layers (or, equivalently, for the phase shifts between the
Josephson oscillations in the different layers).  This allows us to
reduce the original two-dimensional problem to an one-dimensional one. 
Dynamic equations reveal two families of steady-state solutions:
lattices with constant phase shifts between neighboring layers, starting
from zero for a rectangular configurations to $\pi $ for a triangular
configuration, and double-periodic lattices with the average phase shift
$\pi /2$.  We analyze the stability of the regular lattices and show
that they exhibit several kinds of instabilities including the long-wave
length shear instability and the short-wave length instability with
respect to alternating deformations.
A specific lattice structure at given velocity is selected
uniquely by boundary conditions. The numerical investigation shows
that at small velocities the lattice experiences a smooth evolution
of the structure with the phase shift between neighboring layers
decreasing from $\pi $ at zero velocity to smaller values. At a
certain velocity the lattice becomes unstable and the system switches
into an oscillating or chaotic state.
At even higher velocities the steady-state regular lattice is
restored again.  The phase shift for this rapidly moving lattice
is smaller than $\pi /2$ and continues to decrease with increasing
velocity, i.e., the structure evolves smoothly towards a
rectangular configurations.  The current-voltage characteristic is
non-monotonic in the region of the instabilities, which means that
two different coherent steady-states coexist within a certain
current range.  Within this current range the system can also be
in the phase separated states, in which it is split into two (or
more) domains moving with different velocities separated by a
phase defect.

The paper is organized as follows.  In Section \ref{Sec-DynEq} we derive
the coupled dynamic equations for the in-plane fields and phase differences. 
In Section \ref{Sec-EqPhSh} we derive the dynamic equations for the relative
displacements [phase shifts] between the Josephson planar arrays in
layers.  In Section \ref{Sec-CohStSt} we study the coherent
steady-states [regular lattices] and calculate the excess Josephson
currents and Poynting vectors for such states.  In Section
\ref{Sec-Stab} we investigate the stability of the steady-states.  We
consider in detail two important particular cases of instabilities: the
long-wave and short-wave instability.  We also compute numerically the
stability phase diagrams in the plane structure-Josephson frequency.  In
Section \ref{Sec-BoundStruc} we consider boundary conditions for top and
bottom boundaries, discuss the mechanism of structure selection by the
boundaries, and numerically calculate the evolution of the lattice
structure with increase of velocity.  Finally, in Section
\ref{Sec-MulBr} we discuss the  mechanism of multiple branches in the
current-voltage dependence due to the dynamic phase separation:
spontaneous splitting of the system into the rapidly and slowly moving
regions.

\section{Dynamic equations}
\label{Sec-DynEq}

The dynamics of the moving Josephson lattice can be described by coupled
equations for the phase differences and magnetic field. These
equations can be derived from Maxwell's equations expressing
fields and currents in terms of the gauge invariant phase
difference between the layers $\theta _{n}=\phi _{n+1}-\phi
_{n}-\frac{2\pi s}{\Phi _{0}}A_{z}$ and the in-plane
superconducting momentum $p_{n}=\nabla _{x}\phi _{n}-\frac{2\pi
}{\Phi _{0}}A_{x}$ (see, e.g., Ref.\ \onlinecite{ArtJETPL97}).
Consider a layered superconductor in a magnetic field applied
along the layers ($y$ -direction) with transport current flowing
along the $c$-axis ($z$-direction). A local magnetic field $H_{n}$
between the layers $n$ and $n+1$ can be expressed as
\begin{equation}
H_{n}(x)=\frac{\Phi _{0}}{2\pi s}\left( \frac{\partial \theta
_{n}}{\partial x}-p_{n+1}+p_{n}\right) .  \label{Hz}
\end{equation}
The components of electric field can be approximately represented
as
\begin{equation}
E_{x}\approx \frac{\Phi _{0}}{2\pi c}\frac{\partial
p_{n}}{\partial t} ;\;E_{z}\approx \frac{\Phi _{0}}{2\pi
cs}\frac{\partial \theta _{n}}{
\partial t}.  \label{ExEz}
\end{equation}
The components of electric current, $j_{x}$ and $j_{z}$, include
the quasiparticle and superconducting contributions
\begin{eqnarray}
j_{x} &=&\sigma _{ab}\frac{\Phi _{0}}{2\pi c}\frac{\partial
p_{n}}{\partial t
}+\frac{c\Phi _{0}}{8\pi ^{2}\lambda _{ab}^{2}}p_{n},  \label{jx} \\
j_{z} &=&\sigma _{c}\frac{\Phi _{0}}{2\pi cs}\frac{\partial \theta
_{n}}{
\partial t}+j_{J}\sin \theta _{n}.  \label{jz}
\end{eqnarray}
where $\sigma _{ab}$ and $\sigma _{c}$ are the components of the
quasiparticle conductivity, $j_{J}=\frac{c\Phi _{0}}{8\pi
^{2}s\lambda _{c}^{2}}$ is the Josephson current density, and
$\lambda _{ab}$ and $ \lambda _{c}$ are the components of the
London penetration depth. Using the above relations we rewrite the
$z$- and $x$-component of Maxwell's equation $\frac{4\pi
}{c}{\bf j}+\frac{\partial {\bf D}}{\partial t}=\nabla \times {\bf
H}$ as
\begin{eqnarray}
\frac{2\sigma _{c}\Phi _{0}}{c^{2}s}\frac{\partial \theta
_{n}}{\partial t}+ \frac{4\pi }{c}j_{J}\sin \theta
_{n}+\frac{\varepsilon _{c}\Phi _{0}}{2\pi c^{2}s}\frac{\partial
^{2}\theta _{n}}{\partial t^{2}} &=&\frac{\partial
H_{n}}{\partial x},  \label{Maxwz} \\
\frac{2\sigma _{ab}\Phi _{0}}{c^{2}}\frac{\partial p_{n}}{\partial
t}+\frac{ \Phi _{0}}{2\pi \lambda _{ab}^{2}}p_{n}
&=&-\frac{H_{n}-H_{n-1}}{s}. \label{Maxwx}
\end{eqnarray}
In the second equation we replaced $\partial H/dz$ by discrete
derivative $ (H_{n}-H_{n-1})/s$ and neglected the in-plane
displacement current (typical frequencies are assumed to be much
smaller than the in-plane plasma frequency $c/\lambda _{ab}$).
Taking discrete derivative of the second equation and using
relation (\ref{Hz}), we rewrite the second equation in terms of
$\theta _{n}$ and $H_{n}$,
\begin{equation}
\left( \frac{4\pi \sigma _{ab}}{c^{2}}\frac{\partial }{\partial
t}+\frac{1}{ \lambda _{ab}^{2}}\right) \left( \frac{\Phi
_{0}}{2\pi s}\frac{\partial \theta _{n}}{\partial x}-H_{n}\right)
=-\frac{H_{n+1}+H_{n-1}-2H_{n}}{s^{2}}. \label{Maxwx1}
\end{equation}
Equations (\ref{Maxwz}) and (\ref{Maxwx1}) describe the dynamics in
terms of phases $\theta _{n}(x,t)$ and field $H_{n}(x,t)$.  They are
equivalent to equations derived in Refs.\
\onlinecite{BulPRB96,ArtJETPL97}.  In the case of negligible in-plane
dissipation ($\sigma _{ab}=0$) these equations can be reduced to coupled
system of equations for the phases $\theta _{n}$, which were used in
Refs.\ \onlinecite{SakaiJAP93,SakaiPRB94,KleinerPRB94,BulPRB94}. 
Equations (\ref{Maxwz}) and (\ref{Maxwx1}) provide a simplified
description of the phase dynamics.  In real systems this dynamics can
also be influenced by the charging effects \cite{Charging} and the
quasiparticle imbalance \cite{QImballance}.  However, at present there
is no experimental information, which would allow to describe these
effects quantitatively.

To simplify the analysis of dynamic equations (\ref{Maxwz}) and
(\ref{Maxwx1}), we introduce the reduced coordinate $u=x/\gamma s$, time
$\tau =\omega _{p}t$, and magnetic field $h_{n}=H_{n}2\pi \gamma \lambda
^{2}/\Phi _{0}$.  We also introduce the reduced penetration depth
$l=\frac{\lambda _{ab}}{s}$ and the dissipation parameters (reduced
conductivities)\cite{McCumber} $\nu _{c}=\frac{4\pi \sigma
_{c}}{\varepsilon _{c}\omega _{p}}$, $\nu _{ab}=\frac{4\pi \sigma
_{ab}\lambda ^{2}\omega _{p}}{c^{2}}$, which are related as $\nu
_{c}/\nu _{ab}=\sigma _{c}\gamma ^{2}/\sigma _{ab}$.  In terms of these
variables Eqs.  (\ref{Maxwz}) and (\ref{Maxwx1}) acquire a much simpler
form
\begin{eqnarray}
\frac{ \partial ^{2}\theta _{n}}{\partial \tau ^{2}}+\nu
_{c}\frac{\partial \theta _{n}}{\partial \tau }+\sin \theta
_{n}-\frac{\partial h_{n}}{\partial u} &=&0, \label{RedTheta} \\
\left( \nabla _{n}^{2}-\frac{1}{l^{2}}\right) h_{n}+\frac{\partial
\theta _{n}}{\partial u}+\nu _{ab}\frac{\partial }{\partial \tau
}\left( \frac{
\partial \theta _{n}}{\partial u}-\frac{h_{n}}{l^{2}}\right) &=&0,
\label{RedH}
\end{eqnarray}
which we will use in our analysis.  Here $\nabla _{n}^{2}$ denotes
the discrete Laplace operator, $\nabla _{n}^{2}h_{n}\equiv
h_{n+1}+h_{n-1}-2h_{n} $.

In certain special cases the quasiparticle conductivity can be
directly related with the Josephson current.  The most known
example of such relation is the Ambegaokar-Baratoff formula for
conventional Josephson junctions.\cite{AmbBarPRL63}  Latyshev
\emph{et.\ al.} \cite{LatyshevPRL99} have found a similar relation
for BSCCO at small temperatures, $j_{J}=\pi \sigma _{c}\Delta
_{0}/(es)$, with $\Delta _{0}$ being the maximum gap.  This
relation was derived assuming d-wave symmetry of the order
parameter and coherent interlayer tunneling.  Using this relation
we obtain a simple estimate for the c-axis dissipation parameter
$\nu _{c}$ at small $T$,
\[
\nu _{c}\approx\frac{\omega _{p}}{2\pi \Delta _{0}},{\text{at
}}T\rightarrow 0.
\]
Taking parameters, typical for slightly underdoped BSCCO at $T\approx
50$K: $\gamma =500$, $\lambda _{ab}=240$ nm, $s=15$ \AA, $\rho
_{ab}\equiv \sigma _{ab}^{-1}=50$ $\mu \Omega \cdot $cm, and $\rho
_{c}\equiv \sigma _{c}^{-1}=600$ $\Omega \cdot $cm, we obtain estimates
$\nu _{ab}\approx 0.1$ and $\nu _{c}\approx 0.002$, which we will use in
the numerical computations.  As we can see, typically the in-plane
dissipation is much stronger than the c-axis dissipation.  Models, which
take into account only the c-axis dissipation channel, are not suitable
for the quantitative comparison with experiments in BSCCO. Meanings,
definitions, and practical formulas for the dimensionless parameters are
summarized in Table 1.

\section{Equations for  relative displacements [phase shifts] between
the Josephson planar arrays.}
\label{Sec-EqPhSh}

At high magnetic fields $H>\Phi _{0}/(5.5\gamma s^{2})$ ($h>l^{2}/5.5$)
strongly overlapping Josephson vortices fill all the
layers.\cite{DenseJosLat} In this regime the Josephson coupling can be
treated perturbatively.  Without the Josephson coupling (i.e., dropping
$\sin \theta _{n}$ in Eq.~(\ref{RedTheta})) the phases $\theta _{n}$
change in space and time as
\begin{equation}
\theta _{n}^{(0)}(\tau ,u)=\omega _{E}\tau +k_{H}u+\phi _{n}
\label{Phase0}
\end{equation}
where the frequency $\omega _{E}$ is set by the electric field
$E_{z}$, $ \omega _{E}=2\pi csE_{z}/(\Phi _{0}\omega _{p})$ and
the smooth function $ k_{H}(u)$ is connected with the local
magnetic field as $d(k_{H}u)/du=h/l^{2} $.  In general, the local
field $h$ varies in space due to self-field of the transport
current.  In the following we consider a situation when the self-field
of current is much smaller then the applied field and neglect its
contribution to $h$.  In this case $k_{H}(u)$ can be approximated
by a constant, $k_{H}=h/l^{2}\equiv 2\pi H\gamma s^{2}/\Phi
_{0}$.  Therefore Eq.\ (\ref{Phase0}) represents the phase
modulation moving with the velocity $v_{E}=\omega
_{E}/k_{H}=cE_{z}/H$.  The phase shift $\phi _{n}$ characterizes
a relative coordinate of the Josephson lattice in the layer $n$.
Change of the phase shift $\delta \phi _{n}$ corresponds to
the displacement of the lattice at distance $\delta \phi _{n}/k_{H}$.
Without the Josephson interaction the system would be degenerate
with respect to the phase shifts $ \phi _{n}$.  The Josephson
coupling eliminates this degeneracy.  It either leads to the slow
dynamics of the phase shifts $\phi _{n}$ or selects a specific
steady-state structure.  We will derive a closed set of equations
governing the dynamics of the phase shifts $\phi _{n}(\tau ) $.
This can be done using an expansion with respect to the Josephson
currents and averaging out rapidly oscillating degrees of freedom.
Since the compression deformations of the lattice have much higher
stiffness in comparison with the shear deformations, one can
neglect the explicit coordinate dependence of $\phi _{n}$.  In
the first order with respect to the Josephson current, the phases
$\theta _{n}(\tau ,u)$ and field $h_{n}(\tau ,u)$ acquire
the oscillating corrections, $\tilde{\theta}_{n}(\tau ,u)$ and
$\tilde{h} _{n}(\tau ,u)$, so that they can be represented as
\begin{eqnarray}
\theta _{n}(\tau ,u) &=&\omega _{E}\tau +k_{H}u+\phi _{n}(\tau
)+\tilde{\theta}_{n}(\tau ,u), \\
h_{n}(\tau ,u) &=&h(u)+\tilde{h}_{n}(\tau ,u),
\end{eqnarray}
with $\langle d\phi _{n}/d\tau \rangle _{n}=0$. Substituting
$\theta _{n}(\tau ,u)$ into Eq.~(\ref{RedTheta}) and averaging it
with respect to $u$ we obtain the current conservation condition
\begin{equation}
\nu _{c}\frac{d\phi _{n}}{d\tau }+i_{Jn}=i_{J}.  \label{phin}
\end{equation}
Here $i_{Jn}=\langle \sin \theta _{n}\rangle _{u}$ is the local
Josephson current between the layers $n$ and $n+1$ averaged with
respect to the in-plane coordinate $u$ and $i_{J}=\langle \sin
\theta _{n}\rangle \equiv \langle \langle \sin \theta _{n}\rangle
_{u}\rangle _{n}$ is the Josephson current averaged over the layer
index. The last current obeys the macroscopic equation $\nu
_{c}\omega _{E}+i_{J}=\partial h/\partial u$. In the following, we
will assume that $d\phi _{n}/d\tau \ll \omega _{E}$. To obtain a
closed set of equations we have to relate the local currents $i_{Jn}$ with 
the phase shifts $\phi
_{m}$ and their time derivatives $d\phi _{m}/d\tau $.  For this we have to find the
oscillating phases and field and average $\sin \theta _{n}$ with
respect to the rapid space variations. The total Josephson current
$i_{Jn}\equiv i_{Jn}[\phi _{m},d\phi _{m}/d\tau ] $ consists of
the nondissipative ($i_{Jn}^{\prime }$ ) and dissipative ($
i_{Jn}^{\prime \prime }$) components
\begin{equation}
i_{Jn}=i_{Jn}^{\prime }+i_{Jn}^{\prime \prime }.
\label{StatDynJosCurr}
\end{equation}
The nondissipative component $i_{Jn}^{\prime }$ determines
interaction between the phase shifts $\phi _{n}$.  To the lowest
order with respect to $d\phi _{n}/d\tau $ it depends only on
the instantaneous configurations of the phase shifts $\phi _{n}$,
$i_{Jn}^{\prime }\equiv i_{Jn}^{\prime }[\phi _{m}]$.  The
dissipative component $i_{Jn}^{\prime \prime }$ gives an additional
contribution to the dissipation coming from the Josephson coupling,
and it is proportional to $d\phi _{n}/d\tau $ for slow time
variations.  The first term is determined by the first order
oscillating corrections to the phases $\tilde{\theta}_{n}(u)$ and
fields $\tilde{h}_{n}$ which can be found from equations
\begin{mathletters}
\begin{eqnarray}
-\frac{\partial ^{2}\tilde{\theta}_{n}}{\partial \tau ^{2}}-\nu
_{c}\frac{
\partial \tilde{\theta}_{n}}{\partial \tau }+\frac{\partial \tilde{h}_{n}}{
\partial u} &=&\sin \left( \omega _{E}\tau +k_{H}u+\phi _{n}\right),
\label{PhaseEq}\\
\left( \nabla _{n}^{2}-\frac{1}{l^{2}}\right)
\tilde{h}_{n}+\frac{\partial \tilde{\theta}_{n}}{\partial u}+\nu
_{ab}\frac{\partial }{\partial \tau } \left( \frac{\partial
\tilde{\theta}_{n}}{\partial u}-\frac{\tilde{h}_{n}}{
l^{2}}\right)  &=&0. \label{HEq}
\end{eqnarray}
\end{mathletters}
The Josephson term in the first equation acts like a source generating
the electromagnetic wave with the frequency $\omega_{E}$ and in-plane
wave vector $k_{H}$.  To obtain solution of this equation we introduce
the phase and field response functions, $G(n,m)\equiv G(n,m;k,\omega )$
and $H(n,m)\equiv H(n,m;k,\omega )$, defined as solutions of equations
\begin{mathletters}
\begin{eqnarray}
\left( \omega ^{2}-i\nu _{c}\omega \right) G(n,m)+ikH(n,m)
&=&\delta _{n,m},
\label{RespPhase} \\
ik\left( 1+i\nu _{ab}\omega \right) G(n,m)+\left( \nabla _{n}^{2}-
\frac{1+i\nu _{ab}\omega }{l^{2}}\right) H(n,m) &=&0.
\label{RespH}
\end{eqnarray}
\end{mathletters}
These complex functions give the oscillating phase and field in
the $m$-th layer generated by the oscillating current in the $n$-th
layer. For bulk response both functions depend only on the difference
$n-m$ and the solution of Eqs.\ (\ref{RespPhase},b) can be obtained
by Fourier transformation. For the function $G(n-m)$ this gives
\begin{eqnarray}
G(n-m;k,\omega ) &=&\int \frac{dq}{2\pi }\exp \left(
iq(n-m)\right)
{\cal G}(q,k,\omega ),\;  \label{RespFun} \\
{\cal G}(q,k,\omega ) &=&\left( \omega ^{2}-i\nu _{c}\omega -
\frac{k^{2}\left( 1+i\nu _{ab}\omega \right) }{2(1-\cos q)+\left(
1+i\nu _{ab}\omega \right) /l^{2}}\right) ^{-1}.  \nonumber
\end{eqnarray}
This function will play a very important role in the following derivations.  In general,
the parameters $\nu_c$ and $\nu_{ab}$ in the last formula may depend on
frequency due to the frequency dependence of the quasiparticle conductivities. 
The pole of the function ${\cal G}(q,k,\omega )$ gives the spectrum of
the plasma waves and their damping.  The spatial dependence of the
function $G(n-m;k,\omega )$ is determined by the complex wave vector
$q_{+}\equiv q_{+}(\omega ,k)$, which describes propagation and decay of
the plasma wave along the $z$ axis for the given frequency $\omega $ and
in-plane wave vector $k$,
\begin{eqnarray}
G(n;k,\omega ) &=&\frac{\delta _{n}}{\omega^{2}-i\nu _{c}\omega}-
\frac{k^{2}\left( 1+i\nu _{ab}\omega\right) }{\left(
\omega^{2}-i\nu _{c}\omega\right) ^{2}}\frac{\exp
(iq_{+}|n|)}{2i\sin q_{+}},  \label{RespFun_n} \\
\cos q_{+} &=&1-\frac{k^{2}\left( 1+i\nu _{ab}\omega\right) }{
2\left( \omega^{2}-i\nu _{c}\omega\right) }+\frac{1+i\nu
_{ab}\omega}{2l^{2}},\;\text{with }\mathop{\rm Im} [q_{+}]>0.
\label{qplus}
\end{eqnarray}
Now we can represent the solution of Eqs.\ (\ref{PhaseEq},b) for
the oscillating phase $\tilde{\theta}_{n}(\tau ,u)$ in the form
\begin{equation}
\tilde{\theta}_{n}(\tau ,u)={\rm Im}\left[
\sum_{m}G(n-m;k_{H},\omega _{E})\exp \left( i\omega _{E}\tau
+ik_{H}u+i\phi _{m}\right) \right],
\label{OscPhaseGen}
\end{equation}
and calculate the nondissipative current $i_{Jn}^{\prime }[\phi
_{n}]$ in Eq.\ (\ref{phin}):
\begin{eqnarray}
i_{Jn}^{\prime }[\phi _{n}] &=&\langle \tilde{\theta}_{n}(\tau ,u)\cos
\left( \omega _{E}\tau +k_{H}u+\phi _{n}\right) \rangle _{u}
\label{StatAver} \\
&=&\frac{1}{2}\sum_{m}{\rm Im}\left[ G(n-m;\omega _{E})\exp \left(
-i\left( \phi _{n}-\phi _{m}\right) \right) \right].   \nonumber
\end{eqnarray}
This term determines the nonlocal dynamic interactions between
different $\phi _{n}$. As follows from Eq.\ (\ref{RespFun_n}),
the range of interaction is given by the length $L_s(k_H,\omega_E)=({\rm Im}
[q_+(k_H,\omega_E)])^{-1}$. At $\omega _{E}=0$ interactions extend
only over the nearest neighbors. The range of interactions
increases with increase of lattice velocity. When the velocity
exceeds the minimum velocity of propagating electromagnetic wave,
the range of interactions is limited only by dissipation.

\begin{figure}
\epsfxsize=3.4in \epsffile{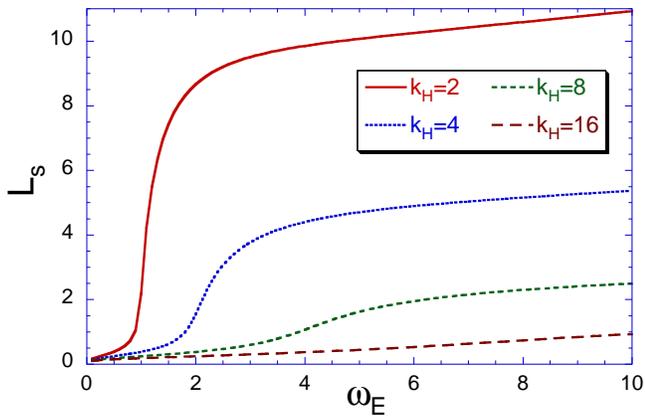} \caption{The frequency dependence of
the decay length $L_s$ for different values of the magnetic wave vector
$k_H$ (see Table 1) for the typical parameters $\nu_{ab}=0.1$ and
$\nu_c=0.002$.  This length determines the interaction range between
the Josephson planar arrays in different layers.  A finite stack,
containing $N$ junctions, shows bulk behavior if $N\gg L_s$.}
\label{Fig-Lzskin}
\end{figure}

To describe the slow phase dynamics, one has to find also the
dissipative component of the Josephson current $i_{Jn}^{\prime \prime
}[\phi _{n},d\phi _{n}/d\tau ]$.  It is determined by the oscillating
contributions to the phases and fields of the order of $\tilde{\theta}
_{n}d\phi _{n}/d\tau $ which we notate as $\theta _{dn}$ and $h_{dn}$. 
Representing $\tilde{\theta}_{n}$ and $\tilde{h}_{n}$ in the complex
form $ \tilde{\theta}_{n},\tilde{h}_{n}\propto \exp \left( i\omega
_{E}\tau +ik_{H}u\right) $, we obtain the following equations for $\theta
_{dn}$ and $ h_{dn}$
\begin{eqnarray*}
\left( \omega _{E}^{2}-i\nu _{c}\omega _{E}\right) \theta
_{dn}+ik_{H}h_{dn} &=&\frac{\partial \phi _{n}}{\partial \tau }\left(
-2\omega _{E}+i\nu _{c}\right) \tilde{\theta}_{n}, \\
\nabla _{n}^{2}h_{dn}-\frac{1+i\nu _{ab}\omega _{E}}{l^{2}}
h_{dn}+ik_{H}\left( 1+i\nu _{ab}\omega _{E}\right) \theta _{dn}
&=&\frac{ \partial \phi _{n}}{\partial \tau }\left( i\nu
_{ab}\frac{\tilde{h}_{n}}{ l^{2}}+\nu
_{ab}k_{H}\tilde{\theta}_{n}\right).
\end{eqnarray*}
Solution for $\theta _{dn}$ can be represented as
\[
\theta _{dn}=\sum_{n_{1},m} \mathop{\rm Im} \left[ F\left(
n-n_{1},n_{1}-m\right) \frac{\partial \phi _{n_{1}}}{\partial \tau
}\exp \left( i\omega _{E}\tau +ik_{H}u+i\phi _{m}\right) \right],
\]
where the kernel $F\left( n,m\right) \equiv F\left( n,m;k_{H},\omega
_{E}\right) $ is defined by relations
\begin{eqnarray}
F\left( n,n_{1}\right)&=&\int \frac{dqdq_{1}}{\left( 2\pi \right)
^{2}}
{\cal F}(q,q_{1})\exp \left(iqn+iq_{1}n_{1}\right) ,\label{DissFun} \\
{\cal F}(q,q_{1}) &=&\left(  \!  -2\omega _{E}+i\nu
_{c}+\frac{ik_{H}^{2}\nu _{ab} }{2(1-\cos q)+\frac{1+i\nu
_{ab}\omega _{E}}{l^{2}}}\frac{2(1-\cos q_{1})}{ 2(1-\cos
q_{1})+\frac{1+i\nu _{ab}\omega _{E}}{l^{2}}} \! \right) {\cal G}(q)
{\cal G}(q_{1}),
\label{DissFunq}
\end{eqnarray}
with ${\cal G}(q)\equiv {\cal G}(q,k,\omega )$ being the response
function defined by Eq.\ (\ref{RespFun}). Using this solution we
can now obtain the term $i_{Jn}^{\prime \prime }[\phi _{n},d\phi
_{n}/d\tau ]$
\begin{eqnarray}
i_{Jn}^{\prime \prime }[\phi _{n},d\phi _{n}/d\tau ] &=&\langle
\theta _{dn}(\tau ,u)\cos \left( \omega _{E}\tau +k_{H}u+\phi
_{n}\right) \rangle _{u}  \label{DynAver} \\
&=&\frac{1}{2} \sum_{n_{1},m} \mathop{\rm Im} \left[ F\left(
n-n_{1},n_{1}-m\right) \frac{\partial \phi _{n_{1}}}{\partial \tau
}\exp \left( i\left( \phi _{m}-\phi _{n}\right) \right) \right].
\nonumber
\end{eqnarray}
Substituting averages (\ref{StatAver}) and (\ref{DynAver}) into
Eq.\ (\ref {phin}) we finally obtain:
\begin{equation}
\sum_{m}\nu _{n-m}\left[ \phi _{j}\right] \frac{d\phi _{m}}{d\tau
}+\frac{1}{ 2}\sum_{m} \mathop{\rm Im} \left[ G(n-m)\exp \left(
i\left( \phi _{m}-\phi _{n}\right) \right) \right] =i_{J},
\label{PhaseDyn}
\end{equation}
where the nonlocal viscosity matrix $\nu _{n-m}\left[ \phi
_{j}\right] $ depends on the instantaneous phase configuration and
is given by
\begin{equation}
\nu _{n-m}\left[ \phi _{j}\right] =\nu _{c}\delta
_{n-m}+\frac{1}{2}\sum_{j} \mathop{\rm Im} \left[ F\left(
n-m,m-j\right) \exp \left( i\left( \phi _{j}-\phi _{n}\right)
\right) \right] \label{ViscGen}
\end{equation}
and the average Josephson current $i_{J}$ is given by
\begin{equation}
i_{J}=\frac{1}{2}\sum_{m}\left\langle \mathop{\rm Im} \left[
G(n-m)\exp \left( -i\left( \phi _{n}-\phi _{m}\right) \right)
\right] \right\rangle _{n,\tau }.
\label{AvJosCur}
\end{equation}
These equations describe the slow dynamics of the phase shifts $\phi
_{n}$ imposed by the Josephson coupling.  Our further analysis is based
on these equations.  Eqs.\ (\ref{PhaseDyn}) and (\ref{ViscGen}) are
valid far away from the top and bottom.  This means that in a stack
consisting of a finite number of junctions $N$, these equation have a
region of validity only if $N$ much larger than the decay length of
electromagnetic wave along the $c$ direction $L_s(k_H,\omega_E)=({\rm
Im}[q_+(k_H,\omega_E)])^{-1}$, which determines the range of the phase
interactions.  In Fig.\ \ref{Fig-Lzskin} we plot the dependence of the
decay length $L_s$ on the Josephson frequency $\omega_E$ for several
values of $k_H$ using parameters $\nu_{ab}=0.1$ and $\nu_c=0.002$.  For
small fields one can see a significant increase of $L_s$ when $\omega_E$
exceeds the minimum frequency, $\omega_{\rm min}=k_H/2$, for the
electromagnetic wave with the in-plane wave vector $k_{H}$. 
Nevertheless for stacks, containing 50-100 junctions, the bulk behavior is
expected in a wide range of fields.  The same stack can be either in the
bulk regime or in the finite-size regime depending on electric and
magnetic fields.

\section{Coherent steady-states and resonant I-V dependencies}
\label{Sec-CohStSt}

\begin{figure}
\epsfxsize=3.4in \epsffile{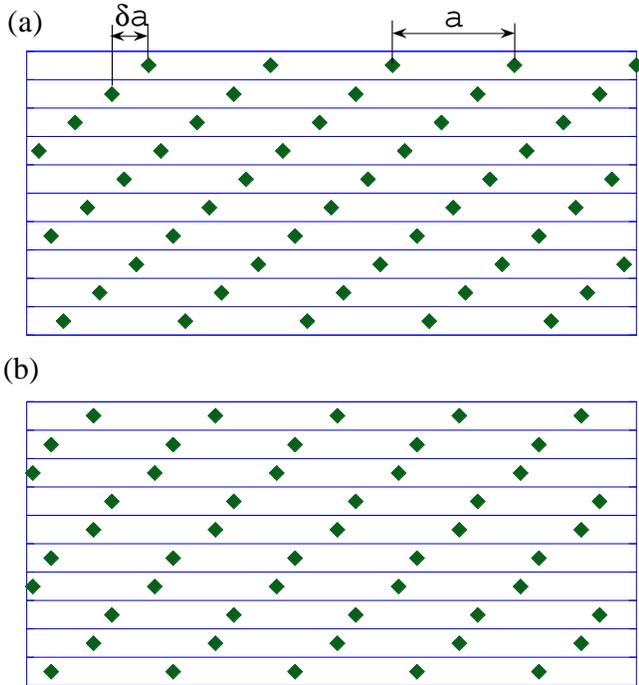} \caption{Steady-states of
moving Josephson lattice (squares mark positions where the interlayer 
phase difference is equal to $\pi+2\pi m$): (a) a periodic lattice, the structure is
determined by the lattice wave vector $\kappa$, which is
connected with the shift between the lattices in the neighboring
layers $\delta a$ by relation $\kappa=2\pi\delta a/a$
(b)a double-periodic lattice }
\label{Fig-StSt}
\end{figure}

In steady-states we have $d\phi _{n}/d\tau =0$ and the static
phase shifts $\phi _{n}$ far away from the boundaries obey equations
\begin{equation}
\frac{1}{2}\sum_{m} \mathop{\rm Im} \left[ G(n-m)\exp \left(
-i\left( \phi _{n}-\phi _{m}\right) \right) \right] =i_{J}.
\label{StState}
\end{equation}
These equations have a trivial solution
\begin{equation}
\phi _{n}=\kappa n,  \label{PerLat}
\end{equation}
corresponding to a regular lattice. In this solution the
planar arrays in the neighboring layers are shifted by
the fraction $\kappa /2\pi $ of the lattice constant (see
Fig.~\ref{Fig-StSt}a).  In particular, $\kappa =0$ corresponds to
a rectangular lattice, and $\kappa =\pi $ corresponds to a triangular
lattice giving a ground state at $\omega _{E}=0$.  The oscillating
phase (\ref{OscPhaseGen}), $\tilde{\theta}_n\equiv
\tilde{\theta}_n(\kappa , k_{H},\omega _{E})$, and  the average
Josephson current (\ref{AvJosCur}), $i_{J}\equiv i_{J}(\kappa ,
k_{H},\omega _{E})$, for the moving regular lattice are given by
\begin{eqnarray}
\tilde{\theta}_n&=&\mathop{\rm Im}\left[ {\cal G}(\kappa
,k_{H},\omega _{E})\exp i\left(\omega_E\tau+ k_H u+\kappa n
\right)\right],
\label{OscPhase}\\
 i_{J} &=&\frac{1}{2}\mathop{\rm Im}\left[ {\cal
G}(\kappa ,k_{H},\omega _{E})\right]. \label{iJ}
\end{eqnarray}
In the case of relatively weak in-plane dissipation, $\nu
_{ab}\omega _{E}\ll 2l^{2}(1-\cos \kappa )+1$, the Josephson
current acquires a simple resonant dependence on the Josephson
frequency $\omega _{E}$ (electric field)
\begin{equation}
i_{J} =\frac{1}{2}\frac{\omega _{E}\nu (\kappa )}{\left( \omega
_{E}^{2}-\omega _{p}^{2}(\kappa )\right) ^{2}+\left( \omega
_{E}\nu (\kappa )\right) ^{2}} \label{iJRes}
\end{equation}
where
\[
\omega _{p}(\kappa )=\frac{k_{H}}{\sqrt{2\left( 1-\cos \kappa
\right) +1/l^{2}}}
\]
is the plasma frequency at the wave vector $\kappa $ and
\[
\nu (\kappa )\equiv \nu _{c}+\frac{2\left( 1-\cos \kappa \right)
k_{H}^{2}\nu _{ab}}{\left( 2\left( 1-\cos \kappa \right)
+\frac{1}{l^{2}} \right) ^{2}}.
\]
is the $\kappa $-dependent dissipation parameter. When the
Josephson frequency $ \omega _{E}$ matches the frequency of the
corresponding plasma wave $\omega _{p}(\kappa )$, a resonance
enhancement of the current is expected. Using this result we can
represent the current-voltage characteristic as
\begin{equation}
j_{z}=\sigma _{c}E_{z}\left( 1+\frac{1}{2}\frac{E_{p}^{4}\nu
(\kappa )/\nu _{c}}{\left( E_{z}^{2}-E_{r}^{2}(\kappa )\right)
^{2}+\left( E_{p}E_{z}\nu (\kappa )\right) ^{2}}\right),
\label{IVa}
\end{equation}
where
\[
E_{r}(\kappa )=\frac{Hs/\lambda _{ab}}{\sqrt{\varepsilon
_{c}}\sqrt{2\left( 1-\cos \kappa \right) +s^{2}/\lambda _{ab}^{2}}}
\]
is the resonance electric field and $E_{p}\equiv \Phi _{0}\omega
_{p}/(2\pi cs)$ is the electric field at which the Josephson
frequency matches the plasma frequency.  This expression is
similar to Eck peak for a single junction.\cite{Eck} However,
in contrast to a single junction, both the resonance frequency and
the dissipation parameter depend on lattice structure.  An
important scale of the electric field is set by the minimum
frequency of the propagating plasma modes, corresponding to the triangular
lattice
\[
E_{\rm{min}}\equiv E_{r}(\pi)\approx {\frac{Hs}{4\lambda
_{ab}\sqrt{\varepsilon _{c}} }}.
\]
For BSCCO ($\lambda_{ab}=220$ nm, and $s= 15$ \AA, $\varepsilon
_{c}=11$) this field typically corresponds to the voltage drop of
$\approx 0.48$ mV per junction at $H = 1$ tesla. At fixed electric
field the triangular configuration ($\kappa =\pi $) gives the
highest dissipation and the highest current, while the rectangular
configuration ($ \kappa =0$) gives the lowest dissipation and the
smallest current.  There are two reasons for this property.
Firstly, the triangular configuration has the smallest resonance
frequency $\omega _{p}(\pi )\approx k_{H}/2$ and the rectangular
configuration has the largest resonance frequency $\omega
_{p}(0)=lk_{H}$ exceeding $\omega _{p}(\pi )$ by the factor
$2l=2\lambda_{ab}/s\approx 300$.  Secondly, the in-plane contribution to
dissipation is completely absent for the rectangular lattice and
it is maximum for the triangular lattice.  Eq.\ (\ref{IVa}) can
be used for the direct comparison with experiment only at small
electric fields, where the structure is close to triangular one.
At higher electric fields the lattice structure (parameter $
\kappa $) changes with the electric field.  This evolution of the lattice
structure is determined by the boundary conditions and will be
considered below.

The motion of the regular Josephson lattice generates a traveling
electromagnetic wave inside the superconductor.  The energy flux in this
wave is given by the Poynting vector
\[
{\bf P}=\frac{c}{4\pi }[{\bf \tilde{E}}\times {\bf \tilde{H}}]
\]
where ${\bf \tilde E}$ and ${\bf \tilde H}$ are the oscillating
components of the electric and magnetic fields.  Using expression
for the $y$ component of the magnetic field, $\tilde{H}_{y}$, and for
the components of the electric field, $\tilde{E}_{x}$ and
$\tilde{E}_{z}$,
\begin{eqnarray*}
\tilde{H}_{y}&=&\frac{\Phi _{0}}{2\pi \gamma \lambda
^{2}}\mathop{\rm Im}\left[\frac{ik_{H}(1+i\nu _{ab}\omega
_{E}){\cal G}(\kappa,k_{H},\omega_{E} )}{2(1-\cos \kappa )+(1+i\nu
_{ab}\omega _{E})/l^{2}} \exp \left( i\left(
k_{H}x+\kappa n+\omega _{E}t\right) \right)\right],\\
\tilde{E}_{x}&=&\frac{\Phi _{0}\omega _{p}}{2\pi cs\gamma
}\mathop{\rm Im} \left[ \frac{k_{H}\omega _{E}\left( 1-\exp
(-i\kappa )\right){\cal G}(\kappa,k_{H},\omega_{E} ) }{2(1-\cos
\kappa ) +(1+i\nu _{ab}\omega _{E})/l^{2}} \exp \left( i\left(
k_{H}x+\kappa n
+\omega _{E}t\right) \right) \right], \\
\tilde{E}_{z} &=&\frac{\Phi _{0}\omega _{p}}{2\pi cs} \mathop{\rm
Im} \left[ i\omega _{E}{\cal G}(\kappa,k_{H},\omega_{E} ) \exp
\left( i\left( k_{H}x+\kappa n+\omega _{E}t\right) \right)
\right],
\end{eqnarray*}
we obtain the components of the Poynting vector,
\begin{eqnarray}
P_{x}=P_{ab}\frac{\omega _{E}k_{H}\left( 2(1-\cos \kappa )+(1-\nu
_{ab}^{2}\omega _{E}^{2})/l^{2}\right) }{\left| 2(1-\cos \kappa
)+(1+i\nu _{ab}\omega _{E})/l^{2}\right| ^{2}}\left|{\cal
G}(\kappa,k_{H},\omega_{E} )\right|^{2}, \label{Poyntingx}\\
P_{z}=P_{c}\frac{\omega _{E}k_{H}^{2}\left( \sin \kappa -\nu _{ab}\omega
_{E}\left( 1-\cos \kappa \right) \right) }{\left| 2(1-\cos \kappa
)+(1+i\nu _{ab}\omega _{E})/l^{2}\right| ^{2}}\left|{\cal
G}(\kappa,k_{H},\omega_{E} )\right|^{2},\label{Poyntingz}
\end{eqnarray}
with
\begin{equation}
P_{ab}=\frac{\Phi _{0}^{2}\omega _{p}}{32\pi ^{3}\lambda
^{2}s\gamma },\; P_{c}=P_{ab}/\gamma. \label{PoyntScale}
\end{equation}
For the typical parameters of BSCCO ($\lambda=200$nm, $\gamma=$500,
$\omega_{p}/2\pi=150$ GHz), the scale of the in-plane Poynting
vector can be estimated as, $P_{ab}\approx 125$ W/cm$^{2}$.

At small electric fields the lattice is triangular and Eq.\
(\ref{IVa}) gives the linear flux-flow conductivity.  We represent
it in the form convenient for comparison with experiment
\begin{equation}
\sigma _{c}^{ff}\approx \sigma _{c}\left( 1+\frac{1}{2}\left(
\frac{\Phi _{0} }{\pi H\gamma s^{2}}\right) ^{4}\right)
+\frac{\sigma _{ab}}{2\gamma ^{2}} \left( \frac{\Phi _{0}}{\pi
H\gamma s^{2}}\right) ^{2} \label{SigmaFF}
\end{equation}
We neglected here small terms $\sim 1/l^{2}$.  This result, valid at
$H>\Phi _{0}/\pi \gamma s^{2}$, was first derived in Ref.\
\onlinecite{BulPRB96}.  An important feature of this dependence is the
$H^{-2}$ term coming from the in-plane dissipation.  In high-T$_{c}$
superconductors suppression of scattering leads to a strong enhancement
of the in-plane quasiparticle conductivity, which is seen as a pronounced
peak in the temperature dependence of $\sigma _{ab}$
\cite{SigmaQYBCO,SigmaQBSCCO}.  On the other hand, $\sigma _{c}$
monotonically decreases when the temperature goes down.  The inequality
$\sigma _{ab}\gg \sigma _{c}\gamma ^{2}$ holds almost everywhere in the
superconducting state.  In this case Eq.\ (\ref{SigmaFF}) predicts
the quadratic dependence of the flux flow resistivity $\rho _{c}^{ff}\propto
H^{2}$ in a wide field range $\Phi _{0}/\pi \gamma s^{2}<H< \sqrt{\sigma
_{ab}/\sigma _{c}\gamma ^{2}}\Phi _{0}/\pi \gamma
s^{2}$.\cite{KoshPRB00} In contrast, for the case of dominating c-axis
dissipation, the flux-flow conductivity at high fields coincides with
the c-axis quasiparticle conductivity and it is field independent.  Eq.\
(\ref{SigmaFF}) describes very well the field dependence of the
flux-flow resistivity in BSCCO.\cite{LatyshSendai}

Another important case, which is realized at high currents, is the
rectangular lattice with $\kappa = 0$.  In this limit Eq.\ (\ref{IVa})
significantly simplifies
\begin{equation}
j_{z}=\sigma _{c}E_{z}\left( 1+\frac{1}{2}\frac{E_{p}^{4}}{\left(
E_{z}^{2}-E_{r}^{2}(0 )\right) ^{2}+\left( E_{p}E_{z}\nu
_{c}\right) ^{2}} \right),  \label{IVhigh}
\end{equation}
where $E_{r}(0 )=H/\sqrt{\varepsilon _{c}}$ is the largest
resonance field.  Note that in this limit the in-plane dissipation
does not influence any more the lattice dynamics.  A pronounced
resonant feature at $E_{z}=E_{r}(0)$ can be observed only for the
case of very small c-axis dissipation.  The peak amplitude is
comparable with the quasiparticle current $\sigma _{c}E_{z}$ if
\[
H\lesssim \Phi_{0}/(2\pi \nu_{c} \lambda_{c}s).
\]
For typical BSCCO parameters this corresponds to fields $\lesssim 1$T.
Another obvious condition for observation of the resonance is that the
voltage drop between the neighboring layers at the resonance field
$sE_{r}(0 )$ must be smaller than the gap voltage $\Delta _{0}/e$, which
gives
\[
H<H_{\Delta }=\frac{\Delta _{0}\sqrt{\varepsilon _{c}}}{es}
\]
Using $\Delta _{0}=25$mV and $\varepsilon _{c}=12$ we obtain
$H_{\Delta }\approx 2$kG.

In addition to the periodic lattices, Eq.\ (\ref{StState}) allows for
the double-periodic solutions of the form
\[
\phi _{n}=\frac{\pi }{2}n+u(-1)^{n}
\]
This can be shown by the direct substitution of this expression into
Eq.\ (\ref {StState}) using the following identities
\begin{eqnarray*}
\exp \left( -iu\left( (-1)^{n}-(-1)^{m}\right) \right) &\equiv
&1+\frac{
\left( 1-(-1)^{n-m}\right) }{2}(\exp \left( -i(-1)^{n}2u\right) -1) \\
\sum_{m}G(m)\exp \left( -i\frac{\pi }{2}m\right) &\equiv
&\sum_{m}G(m)\exp \left( i\frac{\pi }{2}m\right)
\end{eqnarray*}
Such double-periodic lattice is sketched in Fig.~\ref{Fig-StSt}b.
The current $i_{J}$ for these states does not depend on the
modulation parameter $u$, $i_{J}(k_{H},\omega _{E})=\frac{1}{2}
\mathop{\rm Im} \left[ {\cal G}(\frac{\pi }{2},k_{H},\omega
_{E})\right] $.

\section{Stability of coherent steady-states}
\label{Sec-Stab}

\subsection{Equations for small deviations from regular lattice
and stability criterion}

To investigate stability of the moving regular lattice we consider
a solution of the form $\phi _{n}(\tau )=\kappa n+u_{n}(\tau )$ and
obtain equations for the small deformations $u_{n}(\tau )\ll 1$
\begin{eqnarray}
\nu _{c}\frac{\partial u_{n}}{\partial \tau
}+\frac{1}{2}\sum_{n_{1}} \mathop{\rm Im} \left[ F\left(
n-n_{1},\kappa \right) \exp \left( i\kappa (n_{1}-n)\right)
\right] \frac{\partial u_{n_{1}}}{\partial \tau } &&  \label{LinEq} \\
+\frac{1}{2}\sum_{m} \mathop{\rm Re} \left[ G(n-m;\omega _{E})\exp
\left( i\kappa (m-n)\right) \right] \left( u_{m}-u_{n}\right)
&=&0.  \nonumber
\end{eqnarray}
Looking for solutions in the form of plane waves
\[
u_{n}= \mathop{\rm Re} [u_{q}\exp (\alpha (q)\tau +iqn)],
\]
we obtain for the eigenvalue $\alpha (q)\equiv \alpha (q,\kappa,
k_{H},\omega_{E})$
\begin{equation}
\alpha (q)=-\frac{1}{4}\frac{{\cal G}(\kappa +q)+{\cal G}^{\ast
}(\kappa -q)-2 \mathop{\rm Re} \left[ {\cal G}(\kappa )\right]
}{\nu _{c}+\frac{1}{4i}\left[ {\cal F}\left( \kappa +q,\kappa
\right) -{\cal F}^{\ast }\left( -\kappa +q,\kappa \right) \right]
}.  \label{alphaq}
\end{equation}
where the functions ${\cal G}(q)$ and ${\cal F}( q,q_1)$ are given
by Eqs.\ (\ref{RespFun}) and (\ref{DissFunq}). The lattice is
stable if there is no exponentially growing solution in the whole
$q$-interval $0<q<\pi $, i.e.,
\[
\mathop{\rm Re} [\alpha (q)]<0,\;\text{for }0<q<\pi.
\]
The instability is characterized by the wave vector $q_i$, at which
$\mathop{\rm Re} [\alpha (q)]$ becomes positive for the first
time. We consider in detail the two important special cases: the
long-wave instability at $q_i=0$ and the instability with respect
to alternating deformations at $q_i=\pi $.

According to the generally accepted classification (see, e.\ g., Ref.\
\onlinecite{PhysKin}), one can distinguish two kinds of
instability: an absolute instability, for which initial perturbation
exponentially grows at any point and a convective instability, for
which growing perturbation is drifted away so that it decays with
time at fixed point.

\subsection{Long-wave length stability}
\label{Sec-LongWave}

Formally, the condition of the long-wave stability can be obtained from
Eq.\ (\ref{alphaq}) at $q\rightarrow 0$.  However, for better
understanding of the instability mechanism, we will obtain it directly
from the equation of motion for the elastic deformation of the lattice
(\ref{LinEq}).  To derive this equation, we add to the solution
(\ref{PerLat}) slowly changing in space and time function $u(\tau ,n)$,
$\phi _{n}=\kappa n+u(\tau ,n)$, and replace the discrete index $n$ by
a continuous variable $z$.  Substituting this expression into Eq.\
(\ref{PhaseDyn}), expanding it with respect to $u(\tau ,z)$, and
performing a gradient expansion, we obtain equation for the elastic
deformation $u(\tau ,z)$:
\begin{equation}
\nu \frac{\partial u}{\partial \tau }+\zeta \frac{\partial
^{2}u}{\partial z\partial \tau }+a_{1}\frac{\partial u}{\partial
z}+a_{2}\frac{\partial ^{2}u }{\partial z^{2}}+\ldots =0
\label{GradExp}
\end{equation}
with
\begin{mathletters}
\begin{eqnarray}
a_{1} &\equiv &a_{1}\left( \kappa ,k_{H},\omega _{E}\right) =\frac{1}{2}
\mathop{\rm Im} \left[ \frac{\partial {\cal G}(\kappa ,k_{H},\omega
_{E})}{\partial \kappa } \right] , \label{a1}\\
a_{2} &\equiv &a_{2}\left( \kappa ,k_{H},\omega _{E}\right)
=-\frac{1}{4} \mathop{\rm Re} \left[ \frac{\partial ^{2}{\cal
G}(\kappa ,k_{H},\omega _{E})}{\partial
\kappa ^{2}}\right] , \label{a2}\\
\nu &\equiv &\nu \left( \kappa ,k_{H},\omega _{E}\right) =\nu
_{c}+\frac{1}{ 2 } \mathop{\rm Im} \left[ \frac{\partial {\cal
G}(\kappa ,k_{H},\omega _{E})}{\partial \omega _{E}}\right] \equiv
\frac{\partial i(\kappa ,k_{H},\omega _{E})}{\partial
\omega _{E}}, \label{nu}\\
\zeta &\equiv &\zeta \left( \kappa ,k_{H},\omega _{E}\right)
=-\frac{1}{2} \mathop{\rm Re} \left[ \frac{\partial {\cal
F}(\kappa +q,\kappa ,k_{H},\omega _{E})}{
\partial q}\right] _{q=0}.\label{zeta}
\end{eqnarray}
\end{mathletters}
In the derivation of the formula for $\nu $ (\ref{nu}) we used the
relation ${\cal F}(\kappa ,\kappa ,k,\omega )\equiv \partial {\cal
G}(\kappa ,k,\omega )/\partial \omega $.  Note that the
coefficients $a_{1}$ and $\zeta $ vanish for the symmetric lattices
with $\kappa =0,\pi $. Substituting a solution in the form of a plane
wave $u(\tau ,n)\propto \exp \left( \alpha (q)\tau +iqn\right) $
into Eq.\ (\ref{GradExp}) we obtain for small $q$
\begin{equation}
\alpha (q)\approx \frac{1}{\nu }\left( -ia_{1}q+\left(
a_{2}-a_{1}\zeta /\nu \right) q^{2}\right) .  \label{alphaqLong}
\end{equation}
This gives the stability condition $ \mathop{\rm Re} [\alpha (q)]<0$
\begin{equation}
a_{2}-a_{1}\zeta /\nu <0;\;\nu >0.  \label{StabCondLong1}
\end{equation}
The first of these conditions means a positive shear stiffness, while
the second condition is equivalent to the condition of the monotonic I-V
dependence at fixed $\kappa$, $ \partial i/\partial \omega _{E}>0$.  On
the other hand, formally, the lattice is also stable when {\em both}
inequalities are opposite
\begin{equation}
a_{2}-a_{1}\zeta /\nu >0;\;\nu <0  \label{StabCondLong2}
\end{equation}
The case $a_1\ne 0$ corresponds to a convective instability.  This
means that only for symmetric lattices the long-wave instability
is an absolute instability. Below we consider several important
special cases of the long-range instability.

For the important particular case of the triangular lattice one can
find a simple analytical equation for the lowest instability
frequency $\omega _{\Delta }(k_{H})$:
\begin{equation}
\omega _{\Delta }^{2}-\frac{k_{H}^{2}}{4}=-\omega _{\Delta }\left(
\nu _{c}+ \frac{k_{H}^{2}\nu _{ab}}{4}\right) \left( \sqrt{1+\nu
_{ab}^{2}\omega _{\Delta }^{2}}-\nu _{ab}\omega _{\Delta }\right)
\label{StabBounTrian}
\end{equation}
and for the current at the instability point
\[
j_{z}=j_{J}\left( \nu _{c}\omega _{\Delta }+\frac{1}{2\omega
_{\Delta }\left( \nu _{c}+\frac{k_{H}^{2}\nu _{ab}}{4}\right)
}\frac{1}{\left( \sqrt{ 1+\nu _{ab}^{2}\omega _{\Delta }^{2}}-\nu
_{ab}\omega _{\Delta }\right) ^{2}+1}\right)
\]
The triangular lattice is stable at $\omega _{E}<\omega _{\Delta
}(k_{H})$, which means that it always becomes unstable before reaching
the resonance frequency $\omega _{p}(\pi )=k_{H}/2$.  The instability
point approaches the resonance with decreasing dissipation.  In the
limit of high in-plane dissipation $\nu _{ab} \gg 2/k_{H}$ the
instability point $\omega _{\Delta }$ approaches the universal
dissipation-independent value $\omega _{\Delta }\rightarrow
k_{H}/2\sqrt{2}=\omega _{p}(\pi )/\sqrt{2}$

The stability boundary of the rectangular lattice $\omega _{\Box
}(k_{H})$ is given by
\[
\omega _{\Box }^{2}-l^{2}k_{H}^{2}=-\omega _{\Box }\nu _{c}\left(
\nu _{ab}\omega _{\Box }+\sqrt{1+\nu _{ab}^{2}\omega _{\Box
}^{2}}\right)
\]
In contrast to the triangular lattice, the rectangular lattice is stable
at $ \omega _{E}>\omega _{\Box }(k_{H})$ and remains stable at the
resonance $\omega _{E}=lk_{H}$.

\subsection{Short-wave length instability}

To investigate stability of the lattice with respect to alternating
deformations we substitute $\phi _{n}=\kappa n+u(-1)^{n}$ into Eq.\
(\ref {PhaseDyn}).  In the linear approximation with respect to the
modulation $u$, it obeys the following simple equation
\begin{equation}
\nu _{\pi }\frac{du}{d\tau }+a _{\pi}u =0,
\end{equation}
where
\begin{eqnarray}
a _{\pi} & \equiv &a _{\pi}(\kappa,k_{H},\omega_{E})= -\frac{1}{2}
\mathop{\rm Re}\left[ {\cal G}(\kappa )-{\cal G}(\pi -\kappa )\right] \\
\nu _{\pi } & \equiv &\nu _{\pi}(\kappa,k_{H},\omega_{E})=\nu
_{c}+\frac{1}{ 2 }\mathop{\rm Im} \left[ {\cal F}\left( \kappa
-\pi ,\kappa \right) \right]
\end{eqnarray}
are the ``$\pi$-stiffness'' and ``$\pi$-viscosity'' constants. The
stability condition is given by an inequality
\begin{equation}
a _{\pi}/\nu _{\pi }>0  \label{ShortWaveA}
\end{equation}
The ``$\pi$-stiffness'' constant has a symmetry property
$a_{\pi}(\kappa,k_{H},\omega_{E}) =
-a_{\pi}(\pi-\kappa,k_{H},\omega_{E})$.  In particular,
$a_{\pi}(\pi/2,k_{H},\omega_{E}) = 0$, which means that the line $\kappa
=\pi /2$ represents the stability boundary for alternating deformations
at all lattice velocities.  This means that no stability region can
cross the line $\kappa=\pi/2$ and \emph{it is impossible to evolve
continuously from the static triangular lattice to the rapidly moving
rectangular lattice without intersecting the instability boundary}.  At
small velocities the lattices with $\kappa <\pi /2$ are stable.  Above a
certain velocity situation is reversed, the lattices with $ \kappa >\pi /2$
become stable.  This reversal point is determined by the condition $d
\mathop{\rm Re} [G(\kappa )]/d\kappa =0$ at $\kappa =\pi /2$.  Any
stability boundary in the region $0<\kappa <\pi /2$ due to the sign change
of $a _{\pi}(\kappa)$ has the symmetric counterpart in the region $\pi
/2<\kappa <\pi $.

\subsection{Stability phase diagrams}

\begin{figure}
\epsfxsize=3.2in \epsffile{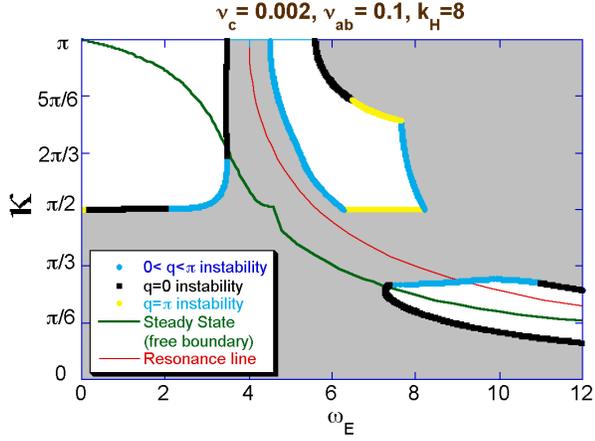}
\caption{Stability regions of the moving Josephson lattice in the plane
$\omega_{E}$-$\kappa$ calculated for the representative parameters
$\nu_{c}=0.002$, $\nu_{ab}=0.1$, and $k_{H}=8$.
Grey regions correspond to unstable lattices.  Sections of boundaries
marked by black correspond to the long-wave instability.  Sections
marked by light grey correspond to the short-wave
instability.  Line starting at ($0$, $\pi$) shows dependence of the
lattice wave vector on the frequency $\omega_{E}$ (lattice velocity)
selected by the ideally reflecting boundary.  We also show the resonance line,
corresponding to matching between the Josephson frequency and the
frequency of the plasma wave at the wave vector $\kappa$.
}
\label{Fig-GenPhaseDiag}
\end{figure}

In general, the lattice can become unstable at an arbitrary wave vector
$0<q<\pi $.  To find the stability regions of the moving lattice, we
numerically scanned the real part of the decay rate $\alpha (q)$
(\ref{alphaq}) throughout the $\omega _{E}-\kappa $ phase diagram and
find boundaries at which $ \mathop{\rm Re} [\alpha (q)]$ changes sign at
least for one $q$.  The obtained stability phase diagram for the
representative parameters $\nu _{c}=0.002 $, $\nu _{ab}=0.1$, and
$k_{H}=8$ is shown in Fig.\ \ref{Fig-GenPhaseDiag}.  For these
parameters we found three stability regions at not very high lattice
velocities (Josephson frequencies): the first low-velocity region is
located below the resonance line and at $\pi /2<\kappa \leq \pi $, the
second one is located above the resonance line and at $\pi /2<\kappa
\leq \pi $, and the third high-velocity region is located along the
resonance line at high velocities with $\kappa $ approaching $0$ with
increase of $\omega _{E}$.  At the boundary of the first region the
lattice experiences the long-wave instability for $\kappa >2.04$.  At
smaller $\kappa $ instability occurs at finite wave vector $q=q_{i}$ and
the instability wave vector $q_{i}$ continuously grows with decrease of
$\kappa $.  We find that this behavior occurs due to the sign change of the
quartic term in the $q$-expansion of $\alpha(q)$ and consider in detail
this transition in Appendix A. The instability at a finite $q$ always 
means that the unstable mode has finite frequency $\omega_{i}$. Close 
to the instability point the system is expected to generate oscillations 
with this frequency.

\begin{figure}
\epsfxsize=3.4in \epsffile{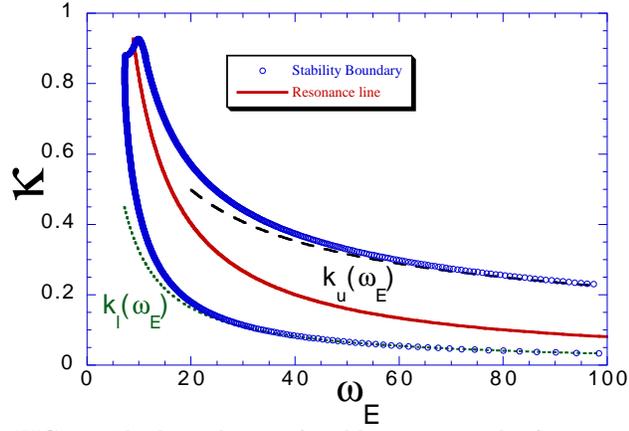} \caption{The
boundaries of stable region in the frequency range $k_{H}\ll \omega
_{E}\ll k_{H}l$ for the same parameter set as in Fig.\
\ref{Fig-GenPhaseDiag}.  The line in the middle is the resonance line. 
Dotted and dashed lines show the asymptotics of the lower and upper
boundaries given by Eqs.\ (\protect\ref{LowerBound}) and
(\protect\ref{UpperBound}).}
\label{Fig-StabHighV}
\end{figure}

In the velocity range $k_{H}\ll \omega _{E}\ll k_{H}l$ there is only one
stability region bounded by two lines above and below the resonance line
$\omega_p(\kappa)\approx k_H/\sqrt{2(1-\cos{\kappa})}$,
$\kappa_l(\omega_E)<\kappa< \kappa_u(\omega_E)$, see Fig.\
\ref{Fig-StabHighV}.  Both lines correspond to the long-wave
instability.  In a wide range of frequencies these lines are described
by simple analytical formulae, which we derive in Appendix B:
\begin{eqnarray}
\kappa _{l}(\omega _{E})&=&\frac{k_{H}}{\sqrt{6}\omega _{E}}, \text{ at
} 1/\nu_{ab}<\omega_{E}< (k_{H}l)^{2/3}/(6\nu_{ab})^{1/3}\label{LowerBound} \\
\kappa _{u}(\omega _{E})&=&k_{H}\sqrt{\sqrt{\frac{3}{5}}\frac{\nu
_{ab}}{ \omega _{E}}},  \text{ at} 1/\nu_{ab}<\omega _{E}< \nu
_{c}^{-1/3} \label{UpperBound}
\end{eqnarray}
As one can see from Fig.\ \ref{Fig-StabHighV}, these asymptotics
agree very well with the numerically calculated stability boundaries.

\begin{figure}
\epsfxsize=3.4in \epsffile{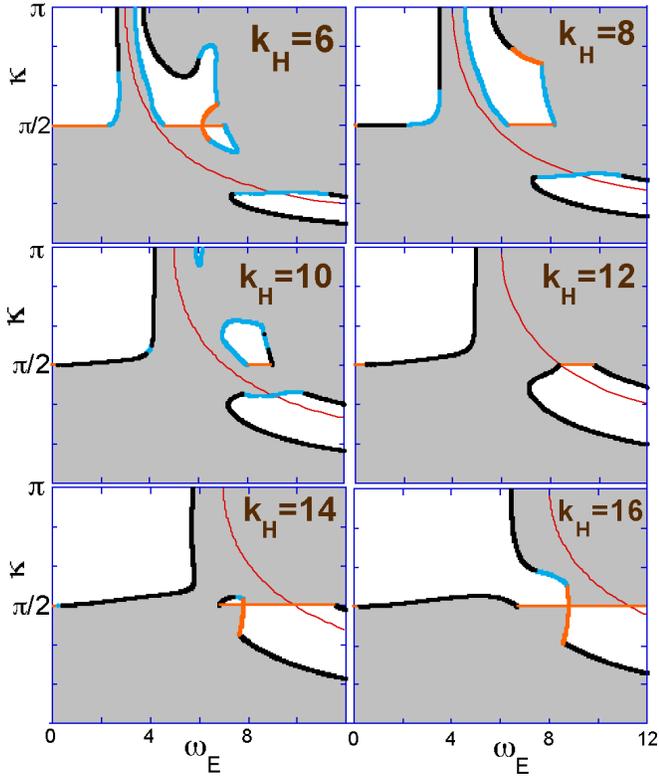}
\caption{Evolution of the stability diagram with increase of the magnetic field
(for $\gamma= 500$ $H\approx k_{H}\cdot 0.3$T).}
\label{Fig-PhaseDiagH}
\end{figure}

Fig.\ \ref{Fig-PhaseDiagH} shows the evolution of the stability
diagram with increase of magnetic field (parameter $k_{H}$, see
Table 1).  Salient features of this evolution are (i) the
stability region above the resonance line shrinks with increase of
field and vanishes at $k_{H}> 10$, (ii) the high velocity region
expands to higher $\kappa$ with increase of field, at $k_H \approx
14 $ additional a small stability island appears above the line
$\kappa =\pi/2$, and at $k_H \approx 16 $ this island merges with
the the low velocity stability region. This roughly corresponds
to the field
$\sqrt{\sigma_{ab}/\sigma_c\gamma^2}\Phi_0/2\pi\gamma s^2$.

\section{Boundary conditions and dynamic structure selection}
\label{Sec-BoundStruc}

\subsection{General case of sharp boundary}
\label{Sec-GenBoun}

Stability analysis of the previous Section does not show what
structure is actually realized at given velocity.  For the static
case the structure is selected by the minimum energy condition.
Such condition is absent in the dynamic case.  In this case a
particular structure can be selected by the boundary conditions.
Influence of the boundary on the moving structure is in turn
determined by the interactions of electromagnetic waves with it.

Consider a semiinfinite stack of junctions with $n=1,2\ldots$\ 
separated by a sharp boundary from a medium with arbitrary
electromagnetic properties. We will obtain boundary conditions for
such system in the case of a steady-state. To derive equations for
the phase shifts $\phi _{n}$ for such system we have to find
solution of the linear equations without Josephson coupling
taking into account boundary conditions. For the plasma wave with
given frequency $\omega $ and wave vector along the layers $k$
the oscillating phases ($\tilde{\theta}_{n}$)  and magnetic fields
($\tilde{h}_{n}$) in the junctions can be represented as
\begin{equation}
\tilde{\theta}_{n},\tilde{h}_{n}\propto \exp (-iq_{+}n)+{\cal
B}\exp (iq_{+}n),\text{ at }n\geq 1  \label{hjunc}
\end{equation}
where $q_{+}\equiv q_{+}(k,\omega )$ is the complex wave vector
given by Eq.\ (\ref{qplus}).  Properties of the boundary are
completely characterized by the complex amplitude of reflected
wave ${\cal B}\equiv {\cal B}(k,\omega )$, which has to be found
by matching the solution (\ref{hjunc}) with electromagnetic
oscillations in the medium at $z<0$. In the case of weak
dissipation at $\omega_E > k_H/2$, the wave number $q_+$ has only
small imaginary part and Eq.\ (\ref{hjunc}) describes the usual case
of reflection of a propagating wave. On the other hand, at $\omega_E
< k_H/2$ the wave number $q_+$ is an almost pure imaginary number and
Eq.\ (\ref{hjunc}) describes reflection of a decaying wave,
which is rarely considered in standard electrodynamics. Note
that, in general, $ {\cal B}(k,\omega )$ can be a complex number
with an arbitrary absolute value.  Only in the simplest case of
vanishing dissipation and propagating wave ($ \mathop{\rm Im}
(q_{+})=0$) the amplitude ${\cal B}(k,\omega )$ determines a
conventional reflection coefficient, $R(k,\omega )=\left| {\cal
B}(k,\omega )\right| ^{2}$, and has property $\left| {\cal
B}(k,\omega )\right| <1$. In the continuum limit, $q_{+}\ll 1$,
${\cal B}(k,\omega )$ is given by Fresnel formula (see, e.g.,
Ref.\ \onlinecite{Fresnel})
\begin{equation}
{\cal B}(k,\omega )= {\frac{\varepsilon q_{+}/s-
\varepsilon_{ab}(\omega) q_t }{\varepsilon q_{+}/s+
\varepsilon_{ab}(\omega) q_t}} \label{Fresnel}
\end{equation}
where $\varepsilon_{ab}(\omega)=\varepsilon _{ab0}+\frac{4\pi
\sigma _{ab}}{i\omega }-\frac{c^{2}}{\lambda _{ab}^{2}\omega
^{2}}$ is the in-plane dielectric constant of the superconductor,
$\varepsilon$ is the dielectric constant of the medium at $z<0$,
and $q_t= \sqrt{\frac{\varepsilon \omega ^{2}}{c^{2}}-k^{2}}$ is
the wave vector of the transmitted wave.

To derive equations for the phase shifts valid close to the
boundary one has to find the phase response function $G(n,m)$ from
Eqs.\ (\ref{RespPhase}, b) with appropriate boundary conditions.
Due to relation (\ref{hjunc}) the functions $G(n,m)$ should behave
as
\[
G(n,m)\propto \exp \left( -iq_{+}(n-m)\right) +{\cal B}\exp
(iq_{+}(n+m)), \text{ at }1\leq n\leq m.
\]
Solution of Eqs.\ (\ref{RespPhase},b) satisfying this condition
can be constructed as
\begin{equation}
G(n,m)=G(n-m)+{\cal B}G(n+m),  \label{RespSurfGen}
\end{equation}
where the second term describes the surface contribution and
vanishes at $ n,m\rightarrow \infty $.   Equations for the
steady-state phase shifts $\phi _{n}$, valid near the boundary,
have the following form
\begin{equation}
\frac{1}{2}\sum_{m=1}^{\infty } \mathop{\rm Im} \left[G(n,m) \exp
\left( -i\left( \phi _{n}-\phi _{m}\right) \right) \right] =i_{J}
\label{StStateBound}
\end{equation}
In general, a simple anzats $\phi _{n}=\kappa n$ does not satisfy
the steady-state equations (\ref{StStateBound}) near the surface.
A general solution has the form $\phi _{n}=\kappa n+u_{n}$, where
$u_{n}$ is the surface deformation, $u_{n}\rightarrow 0$ at
$n\rightarrow \infty $.  Equations for $u_{n}$ can be represented
in the form
\begin{equation}
\frac{1}{2}\sum_{m=1}^{\infty }\mathop{\rm Im}\left[ G(n,m)\exp
\left( -i\kappa (n-m)\right) \left( \exp \left( -i\left(
u_{n}-u_{m}\right) \right) -1\right) \right] =i_{s}(\kappa, n)
\label{SurfDef}
\end{equation}
where
\begin{equation}
i_{s}(\kappa,n)=i_{J}-\frac{1}{2}\sum_{m=1}^{\infty }\mathop{\rm
Im}\left[ G(n,m)\exp \left( -i\kappa (n-m)\right) \right]
\label{SurfCurr}
\end{equation}
is the excess Josephson current near the surface,
$i_{s}(\kappa,n)\rightarrow 0$ at $n\rightarrow \infty $.  The
system (\ref{SurfDef}) is degenerate because it contains only the
differences $u_{n}-u_{m}$.  As a consequence, the solution for
$u_{n}$ exists only for certain values of $\kappa$, i.e.,
\emph{bulk structure is selected by the boundary conditions}. This
mechanism of structure selection is realized in many dynamical
systems (for a general review see Ref.\ \onlinecite{DynStructSel}).

The condition for structure selection can be formulated in the
explicit form in the case of small surface deformations $u_{n}\ll
1$, which is always realized at small velocities. In this case
Eqs.\ (\ref{SurfDef}) are reduced to a linear system
\begin{equation}
\frac{1}{2}\sum_{m=1}^{\infty } \mathop{\rm Re} \left[ G(n,m) \exp
\left( -i\kappa (n-m)\right) \right] (u_{n}-u_{m})=i_{s}(\kappa ,n). 
\label{SurfLin}
\end{equation}
The system (\ref{SurfLin}) is degenerate and its solution exists only
for special values of $\kappa $.\cite{LinearSystem} To formulate
condition for the existence of the solution one has to find a solution
$\psi _{n}$ of the adjoint homogeneous equation
\begin{equation}
\sum_{m=1}^{\infty }\left( \mathop{\rm Re} \left[ G(n,m)\exp
\left( -i\kappa (n-m)\right) \right] \psi _{n}- \mathop{\rm Re}
\left[ G(n,m)\exp i\kappa (n-m)\right] \psi _{m}\right) =0.
\label{AssocHom}
\end{equation}
Eq.\ (\ref{SurfLin}) has a nontrivial solution only if its
right-hand side is orthogonal to $\psi _{n}$, i.e., 
\begin{equation}
\sum_{n=1}^{\infty }\psi _{n}i_{s}(\kappa ,n)=0 . \label{SelCrit}
\end{equation}
Eqs.\ (\ref{AssocHom}) and (\ref{SelCrit}) determine the lattice
wave vector $\kappa $ at small velocities.

For finite system, $n=1,2,\ldots, N$, with identical boundaries
the configuration is typically symmetric with respect to the
midpoint $n=N/2$, i.e.,
\[
\phi _{n}=\left\{
\begin{array}{l}
\kappa n,\text{ at }1\ll n<N/2 \\
\kappa (N-n),\text{ at }1\ll N-n<N/2
\end{array}
\right.
\]
In general, these solutions do not match at $n=N/2$, which means that
they should be separated by a strongly perturbed region (shock). 
Further numerical simulations confirm such structure of steady-states in
finite systems.
%

Large class of boundaries is well described by the ideal reflection
${\cal B}=-1$.  This includes a boundary with an insulator or free space
(see Appendix C).  In this case the response function $G(n,m)$ in Eq.\
(\ref{StStateBound}) acquires the form
\begin{equation}
G(n,m)=G(n-m)-G(n+m). \label{GnmIdRef}
\end{equation}
We also calculate in Appendix D the reflection amplitude ${\cal
B}(k_H, \omega_E)$ for the more complicated but practically
interesting case of the boundary between the static and moving
Josephson lattices. Below we investigate in detail the evolution of
the lattice structure for the ideally reflecting boundary.

\subsection{Lattice structure at small velocities}

At small velocities the lattice structure is close to the static triangular
configuration and the phase shifts can be represented as
\begin{equation}
\phi _{n}(\tau )=\pi n+u(\tau ,n)  \label{smallVexp}
\end{equation}
with $\left| u(\tau ,n+1)-u(\tau ,n)\right| \ll 1$. Firstly, we
find the lattice wave vector at small velocities from
Eqs.(\ref{SurfCurr}), (\ref {AssocHom}), and (\ref{SelCrit}). At
$\omega _{E}\rightarrow 0$ the function $G(n)$ is real and it is
only nonzero at $n=-1,0,1$ with
\[
G(0)\approx -\frac{1}{k_{H}^{2}}\left( 2+\frac{1}{l^{2}}\right)
;\; G(\pm 1)\approx \frac{1}{k_{H}^{2}}.
\]
In this case the solution $\psi _{n}$ of Eq.\ (\ref{AssocHom}) reduces
to a constant and the condition (\ref{SelCrit}) yields
\begin{equation}
\sum_{n=1}^{\infty }i_{s}(\kappa ,n)=0.  \label{CondSmallV}
\end{equation}
For the ideally reflecting boundary, in a linear approximation with respect to $\omega
_{E}$ and $\chi =\pi -\kappa $, the only nonzero
components of $i_{s}(\kappa ,n)$ at $n=1$ and $2$ are:
\[
i_{s}(\kappa ,1)\approx \frac{\chi }{2k_{H}^{2}}-\left( \nu
_{ab}+\frac{2\nu
_{c}(3+1/l^{2})}{k_{H}^{2}}\right) \frac{\omega _{E}}{2k_{H}^{2}}
,\;i_{s}(\kappa ,2)\approx -\frac{\nu _{c}\omega
_{E}}{2k_{H}^{4}}.
\]
Using these expansions we obtain from (\ref{CondSmallV}) the
following relation
\begin{equation}
\chi =\left( \nu _{ab}+\frac{7+2/l^{2}}{k_{H}^{2}}\nu _{c}\right)
\omega _{E}.  \label{SmallVelDef}
\end{equation}
This relation determines the deformation of the lattice at small
velocities for the ideally reflecting boundary.

To obtain a general dynamic equation for weak lattice
deformations, we substitute expansion (\ref{SmallVelDef}) into
Eq.(\ref{PhaseDyn}), replace the discrete variable $n$ by a
continuous variable $z$, and perform a gradient expansion with
respect to $\partial u/\partial z$. This yields the celebrated Burgers
equation for $u(\tau ,z)$:
\[
\nu \frac{\partial u}{\partial \tau }+a_{2}\frac{\partial
^{2}u}{\partial z^{2}}+b_{2}\left( \frac{\partial u}{\partial
z}\right) ^{2}=\delta i_{J}
\]
with
\begin{eqnarray*}
a_{2} &=&-\frac{1}{4} \mathop{\rm Re} \left[ \frac{d^{2}G(\kappa
;0)}{d\kappa ^{2}}\right] _{\kappa =\pi }\approx - \frac{1}{2k_{H}^{2}},
\\
b_{2} &=&\frac{1}{4} \mathop{\rm Im} \left[ \frac{d^{2}G(\kappa ;\omega
_{E})}{d\kappa ^{2}}\right] _{\kappa =\pi }\approx -\left( \frac{8\nu
_{c}}{k_{H}^{2}}+\nu _{ab}\right) \frac{\omega _{E}}{2k_{H}^{2}}, \\
\nu &=&\nu _{c}+\frac{1}{2} \mathop{\rm Im} \left[ \frac{\partial G(\pi
,\omega _{E})}{\partial \omega _{E}}\right] _{\omega _{E}=0}\approx
\left( 1+\frac{8}{k_{H}^{4}}\right) \nu _{c}+\frac{2 }{k_{H}^{2}}\nu
_{ab}, \\
\delta i_{J} &=&i_{J}-\frac{1}{2} \mathop{\rm Im} \left[ G(\pi ,\omega
_{E})\right] \approx i_{J}-\left( \frac{4\nu _{c}}{ k_{H}^{2}}+\nu
_{ab}\right) \frac{2\omega _{E}}{k_{H}^{2}},
\end{eqnarray*}
Consider a steady-state configuration, $\partial u/\partial \tau
=0$. For a finite system with $N$ junctions and identical
boundaries the deformation is symmetric with respect to the center
$z=N/2$. The regions of regular lattices, located at $z<N/2$ and
$z>N/2$, are separated by the transitional region near the center
where the deformation obeys the equation
\[
\alpha \frac{d^{2}u}{dz^{2}}+\left( \frac{du}{dz}\right) ^{2}=\chi ^{2},
\]
with $\alpha =a_{2}/b_{2}\approx \left[ \left( \nu _{ab}+\frac{8\nu
_{c}}{ k_{H}^{2}}\right) \omega _{E}\right] ^{-1}$, $\chi =\pi -\kappa
$, and condition $du/dz\rightarrow \pm \chi $ at $z-N/2\rightarrow \pm
\infty $.  This equation has the analytical solution (see, e.g., Ref.\
\onlinecite{Kuramoto})
\[
u=\alpha \ln \left( \cosh \frac{\chi }{\alpha }\left( z-\frac{N}{2}
\right) \right).
\]
Using the result (\ref{SmallVelDef}), we obtain that the width of
the transition region $L_{t}=\alpha /\chi $ is given by
\[
L_{t}=\frac{1}{\left( \nu _{ab}+\frac{8\nu _{c}}{k_{H}^{2}}\right)
\left( \nu _{ab}+\frac{7\nu _{c}}{k_{H}^{2}}\right) \omega _{E}^{2}}.
\]
This length diverges at $\omega _{E}\rightarrow 0$ as $\omega
_{E}^{-2}$.  This means that for a finite system there is a typical
crossover frequency $ \omega _{N}$, which is determined by the condition
$L_{t}(\omega _{N})=N$.  At $\omega _{E}<\omega _{N}$ the deformation
field has a parabolic shape, typical for strained static systems.  At
$\omega _{E}>\omega _{N}$ the system is split into the two homogeneously
tilted lattices separated by the shock.

\subsection{Numerical exploration of steady-states and structure evolution}

\begin{figure}
\epsfxsize=3.8in \epsffile{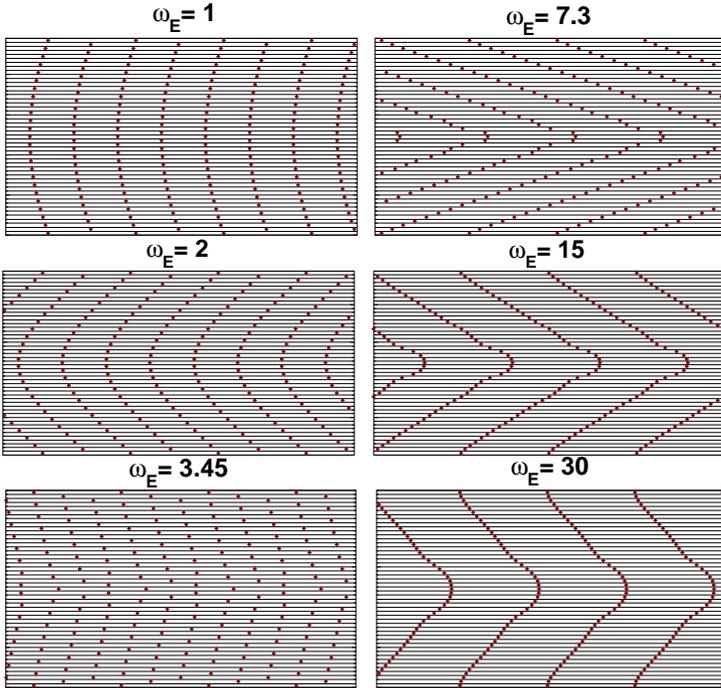} \caption{Lattice
structures in the real space at different values of the Josephson
frequency $\omega_{E}$ (lattice velocity). The structures are
generated using numerically calculated phase shifts for the system
with 51 layers and the same parameters as in
Fig.~\protect\ref{Fig-GenPhaseDiag}. ($\nu_{c}=0.002$,
$\nu_{ab}=0.1$, and $k_{H}=8$). The left column shows the lattice
structures in the low-velocity stability region, below the first
instability point. The right column shows the lattice structures in
the high-velocity stability region. } \label{Fig-LatStruc}
\end{figure}

To find steady-state configurations at all velocities we solved
Eq.\ (\ref{StStateBound}) numerically for the case of ideally
reflecting boundary using the same representative parameters
($\nu_{c}=0.002$, $ \nu_{ab}=0.1$).
The evolution of dependencies $\kappa (\omega _{E})$ obtained from these
solutions for $k_{H}=8$ is shown in Fig.~\ref{Fig-GenPhaseDiag} together
with the stability regions.  We found that at small velocities
(Josephson frequency $\omega _{E}$) the lattice structure smoothly
evolves with increase of velocity and the lattice wave vector $\kappa $
decreases from $\pi $ at zero velocity towards $\pi $/2.  The lattice
structures in this region are shown in the left columns of Fig.\
\ref{Fig-LatStruc}.
At certain velocity the lattice crosses the instability boundary.  This
velocity corresponds to the endpoint of the first flux-flow branch.  It
is close but not identical to the minimum velocity, $\omega_E/k_H=c_{\rm
min}$, of the plasma wave (in reduced units $c_{\rm min}=1/2$ and in 
real units $c_{\rm min}=cs/(2\lambda_{ab}\sqrt{\varepsilon_{c}})\approx 
3\cdot10^{7}$cm/sec).  At
higher velocities the periodic lattice with the phase shift smaller than
$\pi $/2 is restored.  The structure continues to evolve smoothly
towards a rectangular configuration with increasing velocity (see
right column at Fig.\ \ref{Fig-LatStruc}).  

\begin{figure}
\epsfxsize=3.2in \epsffile{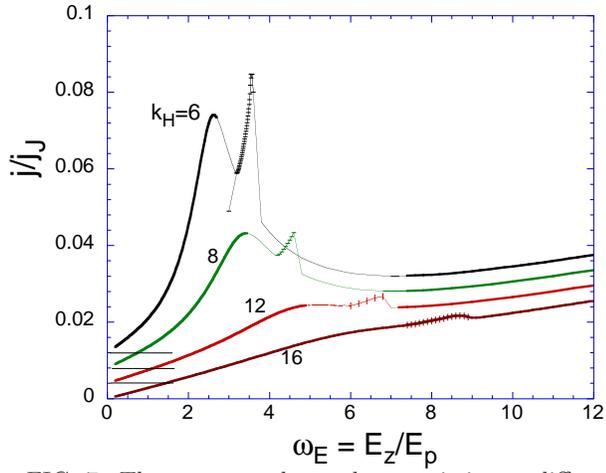}
\caption{The current-voltage characteristics at different magnetic fields,
for $k_{H}=6$, $8$, $12$, and $16$.  The dependencies are obtained using
numerically computed steady-states with parameters $\nu_{c}=0.002$,
$\nu_{ab}=0.1$.  Thick lines show the stable branches and thin lines show
the unstable branches.  The branches, corresponding to the double-periodic
states, are marked by dashes.  The plots are displaced vertically for
clarity.}
\label{Fig-IV}
\end{figure}

Fig.\ \ref{Fig-IV} shows the evolution of the current-voltage
dependence with increase of the magnetic field.  At not very high
magnetic field ($k_{H}<16$) the current-voltage dependencies have
two stable branches, corresponding to the moving regular lattices. The
low-velocity branch corresponds to the structure close to a triangular lattice. It
terminates at the critical velocity. At small magnetic fields one
can see a pronounced current enhancement prior the instability
point. The high-velocity branch corresponds to the structure close to a rectangular
lattice. The resistivity at this branch is close to the $c$-axis
quasiparticle resistivity. In a wide range of magnetic fields, the
stable branches are separated by the broad unstable region where
a homogeneous regular lattice state can not exist.  In this regime
there is a broad range of currents, within which two lattice
solutions exist.

\begin{figure}
\epsfxsize=3.2in \epsffile{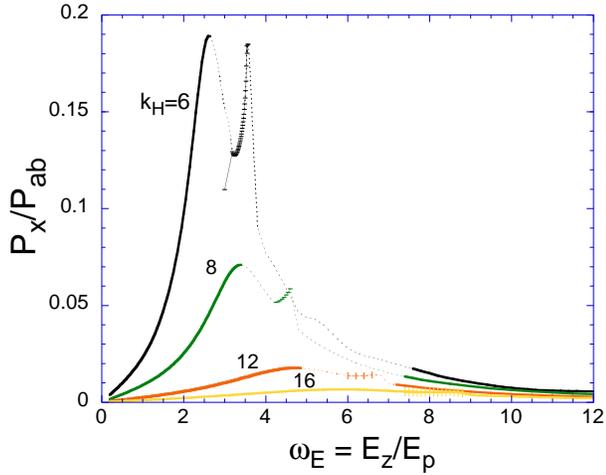} \caption{Electric field
dependencies of the Poynting vector along the layers for electromagnetic
wave generated by the moving lattice (see Eq.\ (\protect
\ref{Poyntingx})).  The scale $P_{ab}$ is defined by Eq.\ (\protect
\ref{PoyntScale}).  We use the same notations for different branches as
in the previous figure.}
\label{Fig-Poyntingx}
\end{figure}

To characterize intensity of the electromagnetic wave generated by the
moving lattice we plot in Fig.\ \ref{Fig-Poyntingx} the electric field
dependencies of the Poynting vector along the layers (\ref{Poyntingx})
at different magnetic fields.  The {\it z} component of the Poynting
vector is always smaller than the {\it ab} component, and its sign
corresponds to the energy flux from the surface to the bulk.  One can
see that the intensity of the electromagnetic wave rapidly increases
with the electric field reaching a maximum at the instability point. 
This enhancement of the intensity of the traveling wave indicates that
the increase of the current near the instability point is caused by the
pumping of energy from a {\emph dc} source into this wave.  This maximum
becomes smaller at higher magnetic fields.  At higher electric fields
the wave intensity decreases with field.  Note however that one can
expect another resonance at very high lattice velocities, where the
velocity reaches the \emph{maximum} velocity of electromagnetic wave,
$c_{\rm max}=c/\varepsilon_{c}$ (see Section \ref{Sec-CohStSt}).  We do
not consider this resonance here.

To clarify the role of specific boundary on the forming the lattice
structure and the current-voltage dependence, we also made several
calculations for the more realistic case of the boundary with the static
lattice.  The amplitude of the reflected electromagnetic wave ${\cal
B}(k, \omega)$ for this case is calculated in Appendix D. Fig.\
\ref{Fig-Compar} shows comparison of the structure evolution and the I-V
dependence for the representative parameters $\nu_{c}=0.002$,
$\nu_{ab}=0.1$, and $k_{H}=6$ with the case of the ideally reflecting
boundary.  Both cases show an overall similar behavior.  The only
qualitative difference exists at very small $\omega_{E}$, where the the
frequency falls within the range of the acoustic branch in the
oscillation spectrum of the static Josephson lattice, $0<\omega
<\sqrt{2}/k_{H}$ (see Appendix D).  In this range the lattice wave
vector exceeds $\pi$ and the energy flux is directed from the bulk to
the surface.  However, this anomaly is almost invisible in the I-V
dependence.  It only becomes noticeable at smaller fields or weaker
dissipation.  In the case of the boundary with the static lattice the
structure is closer to a triangular lattice at the first instability
point.  As a consequence, it has higher current at this point.  This
difference decreases at higher fields.

\section{Multiple branches due to dynamic phase separation}
\label{Sec-MulBr}

\begin{figure}
\epsfxsize=3.3in \epsffile{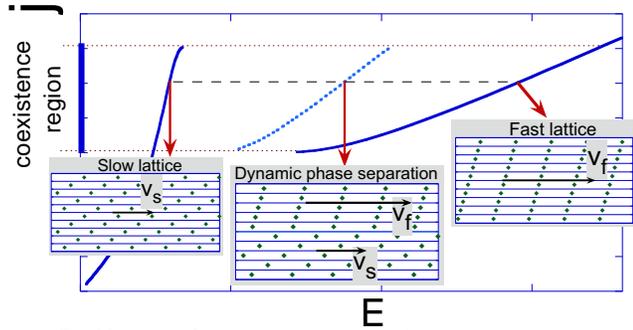} \caption{Multibranch
structure of the current-voltage characteristic due to the dynamic phase
separation.  Two states corresponding of the slow lattice motion (velocity
$v_{s}$) and the fast lattice motion (velocity $v_{f}$) coexist within the
current range marked at the vertical axis.  In this region the
intermediate phase-separated states exist, in which the system is split
into the rapidly and slowly moving regions.  The intermediate branch
corresponding to one of such states is shown by dotted line.}
\label{Fig-PhaseSep}
\end{figure}

We found that for not very high magnetic fields two stable states moving
with different velocities may coexist within a certain range of applied
current.  The dominating in-plane dissipation strongly facilitates such
coexistence, because in this case the lattice velocity is very sensitive
to the lattice structure.  Within the coexistence region one can expect
a family of intermediate states, in which the system is split into the
two regions moving with different velocities separated by a phase
defect (dynamic phase separation, see Fig.\ (\ref{Fig-PhaseSep})).  In a
continuous system coexistence of the states moving with different
velocity at the fixed average velocity is possible only at one driving
force, at which these states are in the dynamic equilibrium.  Out of
equilibrium the boundary will move and eliminate one of the states. 
However, for our discrete system the boundary separating different lattices is
pinned by the discrete structure, and coexistence is possible in a wide
range of driving forces (transport currents).

If the fraction $f_{s}$ of the system is in the slowly moving
state with velocity $v_{s}$ and the remaining fraction $1-f_{s}$
is in the rapidly moving state with velocity $v_{f}$ than the
average velocity $v_{av}$, which determines the observable
electric field, is simply given by
\[
v_{av}=f_{s}v_{s}+(1-f_{s})v_{f}
\]
To investigate the structure of the boundary region, consider an
infinite stack, in which the rapidly moving state occupies the region
$n>0$ and the slowly moving state occupies the region $n<0$.
The condition for coexistence of such states is given by
\[
\nu _{c}\omega _{f}+i_{J}(\omega _{f},\kappa _{f})=\nu _{c}\omega
_{s}+i_{J}(\omega _{s},\kappa _{s})
\]
where $\omega _{f}=k_{H}v_{f}$ ($\omega _{s}=k_{H}v_{s}$) is the
Josephson frequency for the fast (slow) state. In the zero order
approximation with respect to the Josephson coupling the phases
are given by
\begin{eqnarray*}
\theta _{n} &=&\omega _{f}\tau +k_{H}u+\kappa _{f}n+u_{n},\text{ at }n>0 \\
&=&\omega _{s}\tau +k_{H}u+\kappa _{s}n+v_{n},\text{ at }n\leq 0
\end{eqnarray*}
To obtain equations for the boundary deformations $u_{n}$ and
$v_{n}$ we have to find oscillating phases induced by the
Josephson currents. The phases and fields, oscillating with the
frequency $\omega_{f}$, are determined by equations
\begin{eqnarray*}
\nu _{c}\frac{\partial \tilde{\theta} _{n}}{\partial \tau
}+\frac{\partial ^{2}\tilde{\theta} _{n}}{\partial \tau
^{2}}-\frac{\partial \tilde{h}_{n}}{\partial u} &=&-\Theta (n)\sin
\left( \omega _{f}\tau +k_{H}u+\kappa _{f}n+u_{n}\right)
\\
\left( \nabla_{n}^{2}-\frac{1}{l^{2}}\right)
\tilde{h}_{n}+\frac{\partial \tilde{\theta} _{n}}{\partial u}+\nu
_{ab}\frac{\partial }{\partial \tau }\left( \frac{
\partial \tilde{\theta} _{n}}{\partial u}-\frac{\tilde{h}_{n}}{l^{2}}\right)  &=&0
\end{eqnarray*}
where $\Theta (n)$ is the step function. Solution for the oscillating
phase is given by
\[
\tilde{\theta} _{n} ={\rm Im}\left [\sum_{m=0}^{\infty }
G_{f}(n-m) \exp \left( i\left(
 \omega _{f}\tau +k_{H}u+\kappa _{f}n+u_{n}\right) \right)\right ]
\]
with $G_{f}(n-m)\equiv G(n-m; \kappa_{f},\omega_{f})$. Equation
for $u_{n}$ is obtained from condition $\left\langle \sin \theta
_{n}\right\rangle =i_{J}$,
\[
\frac{1}{2}\sum_{m=0}^{\infty } \mathop{\rm Im} [G_{f}(n-m)\exp
\left( i\left( \kappa _{f}(m-n)+u_{m}-u_{n}\right) \right) ]=i_{J}
\]
A similar equation can be derived for the boundary deformations of the
slow state $v_{n}$.  Comparing this equation with Eqs.\
(\ref{RespSurfGen}) and (\ref{StStateBound}), one can see that
the deformations at the boundary between different moving states coincide
with the deformations near the nonreflecting boundary (${\cal B}=0$).

The phase-separated states give the most natural interpretation of
the multiple I-V branches observed by Hechtfischer {\it et al.}
who studied transport properties of Josephson lattice in BSCCO
mesas at high magnetic fields.\cite {HechPRL97} Note that these
branched should not be mixed with the multiple branches at zero
magnetic field, which appear due to the switching of the separate
junctions into the resistive state. This interpretation can be
verified by measuring the spectrum of the microwave irradiation emitting
by the stack. Instead of a single peak located at the Josephson
frequency corresponding to the average voltage, the spectrum of
irradiation should contain two peaks: at the ``slow'' frequency
$\omega _{s}$ with the weight $f_{s}$ and at the ``fast''
frequency $\omega _{f}$ with the weight $1-f_{s}$.

\section{Conclusions}

In summary, we performed a detailed investigation of the Josephson
lattice dynamics in layered superconductors at high fields.  Our
main results can be summarized as follows:
\begin{itemize}
%
\item Interaction between Josephson planar arrays in layers is
mediated by exited electromagnetic oscillations. The dynamics of these
arrays is described by nonlocal and nonlinear equations.

\item For the coherent steady-states, the excess Josephson
current and the Poynting vector of the generated electromagnetic
wave have resonance dependence on the Josephson frequency
(electric field). The resonance frequency is the plasma frequency
at the wave vector selected by lattice structure. In the case of
the dominating in-plane dissipation, typical for high-temperature
superconductors, the damping parameter also strongly depends on
the lattice structure.

\item We investigated stability of the coherent steady-states and found
the two major stability regions in the plane lattice structure-lattice
velocity: the low-velocity region and the high-velocity region.  Exact topology
of the stability phase diagrams depends on dissipation and magnetic field.

\item Lattice structure at given velocity is selected by boundary.
Boundary conditions are determined by the reflection properties of
electromagnetic waves, generated by the moving lattice.

\item We investigated structure evolution with increase of the lattice
velocity. In a wide range of fields there are two stable branches,
the low-velocity branch and the high-velocity branch. At low
velocities the lattice structure is close to a triangular lattice. This
low-velocity branch terminates due to instability at the critical
velocity near the minimum velocity of electromagnetic wave. At
high velocities structure of the lattice approaches a rectangular
configuration.

\item Experimentally observed multiple branches in I-V dependencies
are interpreted as the phase separated states, in which system is
split into the slowly and rapidly moving regions.
\end{itemize}

\section{Acknowledgements}

We thank M.\ Tachiki, R.\ Kleiner,  M.\ Machida, N.\ F.\ Pedersen, Y.\ Latyshev,
and K.\ Gray for helpful discussions.  This work was supported
by the US DOE, Office of Sciences, under contract No.\
W-31-109-ENG-38. AEK also would like to acknowledge support from
the Japan Science and Technology Corporation and to thank the
National Research Institute for Metals for hospitality.

\bigskip
\appendix {\bf Appendix A: Continuous transition from the long-wave
instability to the finite-q instability}
\medskip

\begin{figure}
\epsfxsize=3.3in \epsffile{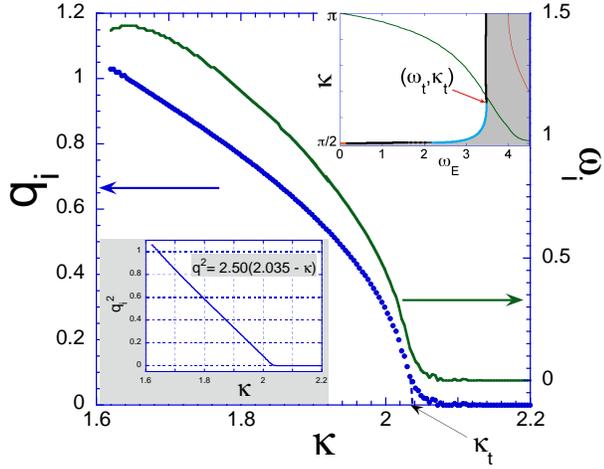} \caption{Behavior of
the wave vector $q_{i}$ (left axis) and frequency
$\omega_{i}=\rm{Im}[\alpha(q_{i})] $ (right axis) of the unstable mode. 
The upper-right inset shows the transition point in the
$\kappa$-$\omega_{E}$ diagram of Fig.\ \protect\ref{Fig-GenPhaseDiag}. 
Lower-left inset shows plot $q_{i}^{2}(\kappa) $ with linear fit below
the transition.}
\label{Fig-Trans}
\end{figure}

Numerically investigating stability of steady-state we found that
the nature of instability of the first flux flow branch depends on
the wave vector $\kappa $, at which the lattice becomes
unstable.  If $\kappa $ is larger than the critical value $\kappa
_{t}$ than instability occurs at $q=0$, i.e., $\partial
^{2}\alpha _{1}/\partial q^{2}$ changes sign.  At $\kappa <\kappa
_{t}$ the lattice becomes unstable at the finite wave vector
$q_{i}$, i.e., $\alpha _{1}(q_{i})=0$ at the instability point.
The transition is continuous, $q_{i}$ smoothly grows with decrease of
$\kappa $ starting from $q_{i}=0$ at $\kappa =\kappa _{t}$ (see
Fig.  \ \ref{Fig-Trans}). This behavior can be explained as
follows.  At small $q$ the decay rate $ \alpha _{1}(q)$ can be
expanded with respect to $q$
\[
\alpha (q)\approx
-a_{2}q^{2}-\frac{1}{2}a_{4}q^{4}-\frac{1}{3}a_{6}q^{6}.
\]
The coefficients $a_{2n}$ in this expansion depend on $\kappa $
and $\omega _{E}$.  For stable lattices $a_{2}>0$.  The
coefficient $a_{2}$ becomes negative when the frequency $\omega
_{E}$ exceeds the long-wave instability boundary $ \omega
_{2}(\kappa )$.  Near $\omega _{2}(\kappa )$ one can use linear
expansion for $a_{2}$
\[
a_{2}=a_{20}(\omega _{2}(\kappa )-\omega _{E}).
\]
Lets also define the frequency $\omega _{4}(\kappa )$, at which
the coefficient $a_{4}$ changes sign. Near $\omega _{4}(\kappa )$
we have
\[
a_{4}=a_{40}(\omega _{4}(\kappa )-\omega _{E}).
\]
If $\omega _{4}(\kappa )>\omega _{2}(\kappa )$ then the
instability occurs at $q=0$. For the first flux flow branch this is
the case for $\kappa >\kappa _{t}$. Otherwise the lattice become
unstable at a finite wave vector $q_{i}$. The transition between these
regimes takes place at the lattice wave vector $\kappa _{t}$
defined by
\[
\omega _{2}(\kappa _{t})=\omega _{4}(\kappa _{t}),
\]
i.e., both $a_{2}$ and $a_{4}$ vanish at $\kappa=\kappa _{t}$.  Lets consider in
detail the behavior in the region where $\kappa $ only slightly smaller
than $\kappa _{t}$ and the difference $\omega _{2}(\kappa _{t})-\omega
_{4}(\kappa _{t})$ is small.  In this region we can use linear
expansions for both $a_{2}$ and $a_{4}$.  We define
\begin{eqnarray*}
\delta &\equiv &\omega _{E}-\omega _{4}(\kappa ), \\
\delta _{2} &=&\omega _{2}(\kappa )-\omega _{4}(\kappa )=\left(
\omega _{2}^{\prime }-\omega _{4}^{\prime }\right) (\kappa -\kappa
_{t})>0,
\end{eqnarray*}
and represent $\alpha (q)$ in the form
\[
\alpha (q)=-a_{20}(\delta _{2}-\delta
)q^{2}+\frac{1}{2}a_{40}\delta q^{4}- \frac{1}{3}a_{6}q^{6}.
\]
With increase of $\delta $ the dependence $\alpha (q)$ becomes
nonmonotonic at $\left( a_{40}\delta \right)
^{2}>4a_{20}a_{6}(\delta _{2}-\delta )$. Extremum values of $q$
are determined by solutions of quadratic with respect to $q^{2}$
equation
\begin{equation}
a_{6}q^{4}-a_{40}\delta q^{2}+a_{20}(\delta _{2}-\delta )=0
\label{Extremum}
\end{equation}
and are given by
\[
q_{\pm }^{2}=\frac{a_{40}\delta }{2a_{6}}\pm \sqrt{\left(
\frac{a_{40}\delta }{2a_{6}}\right) ^{2}-\frac{a_{20}(\delta
_{2}-\delta )}{a_{6}}},
\]
where $q_{-}^{2}$ gives the minimum of $\alpha (q)$ and
$q_{+}^{2}$ gives its maximum. The lattice becomes unstable when
$\alpha (q_{+})=0$. This equation together with the extremum
condition (\ref{Extremum}) determines the instability point
$\delta =\delta _{i}$ and the wave vector of the unstable mode $q_{i}$
\begin{eqnarray*}
\delta _{i} &=&\frac{8}{3}\frac{a_{6}a_{20}}{a_{40}^{2}}\left(
\sqrt{1+\frac{
3a_{40}^{2}\delta _{2}}{4a_{6}a_{20}}}-1\right) , \\
q_{i}^{2} &=&\frac{2a_{20}}{a_{40}}\left(
\sqrt{1+\frac{3a_{40}^{2}\delta _{2}}{4a_{6}a_{20}}}-1\right) .
\end{eqnarray*}
Close to the transition $\delta _{2}\ll a_{6}a_{20}/a_{40}^{2}$ we
have
\begin{eqnarray*}
\delta _{i} &\approx &\delta
_{2}-\frac{3}{16}\frac{a_{40}^{2}}{a_{6}a_{20}}
\delta _{2}^{2}, \\
q_{i}^{2} &\approx &\frac{3a_{40}\delta _{2}}{4a_{6}}.
\end{eqnarray*}
This means that the wave vector $q_{i}$ of the unstable mode
grows as $ q_{i}\propto \sqrt{\kappa _{t}-\kappa }$ near the
transition. The calculated dependence $q_{i}(\kappa )$ agrees very
well with this predictions (see the lower-left inset in Fig.\
\ref{Fig-Trans}).

\bigskip
\appendix {\bf Appendix B: Stability boundaries in the intermediate frequency range}
\medskip

In this appendix we consider the stability region in the
intermediate velocity range, $k_{H}\ll \omega _{E}\ll k_{H}l$,
where it is possible to derive simple analytical expressions for
the stability boundaries. There is only one stability region in
this frequency range located at $\kappa \ll 1$ and limited by two
stability boundaries, $\kappa _{l}(\omega _{E})$ and $\kappa
_{u}(\omega _{E})$, located below and above the resonance line.
Both boundaries $\kappa _{l}(\omega _{E})$ and $\kappa _{u}(\omega
_{E})$ correspond to the long-wave instabilities and we will use
general relations from Section \ref{Sec-LongWave}. Assuming
inequalities $1/l^{2},\nu _{ab}\omega _{E}/l^{2}\ll \kappa ^{2}\ll
1 $, $\nu _{c}\kappa ^{2}\ll \nu _{ab}$ and introducing scaling
variables
\[
z=\kappa ^{2}\omega _{E}^{2}/k_{H}^{2},\;y=\nu _{ab}\omega _{E},
\]
we present the coefficients from Eqs.\ (\ref{a1}-d) in the form
\begin{mathletters}
\begin{eqnarray}
a_{1} &=&-\frac{\kappa }{k_{H}^{2}} \mathop{\rm Im}
\left[ \frac{\left( 1+iy\right) }{\left( z-1-iy\right) ^{2}}\right] ,\label{a1a} \\
a_{2} &=&-\frac{1}{2k_{H}^{2}} \mathop{\rm Re} \left[ \frac{\left(
1+iy\right) \left( 3z+1+iy\right) }{\left( z-1-iy\right)
^{3}}\right] , \label{a2a}\\
\nu  &=&\nu _{c}k_{H}^{3}+\frac{\kappa ^{2}}{2\omega _{E}k_{H}}
\mathop{\rm Im}
\left[ \frac{-2z+iy}{\left( z-1-iy\right) ^{2}}\right] , \label{nua}\\
\zeta  &=&-\omega _{E}k_{H}\kappa ^{3} \mathop{\rm Re} \left[
\frac{2+iy}{\left( z-1-iy\right) ^{3}}\right].\label{zetaa}
\end{eqnarray}
\end{mathletters}
In general, the stability boundaries are determined by Eq.\
(\ref{StabCondLong1}). However, at high enough Josephson frequency
term $a_{1}\zeta /\nu $ can be neglected. We focus here on the
regime $y\gg 1$. Consider the lower boundary $\kappa _{l}(\omega
_{E})$ first. As can be checked from the result for this boundary
the term $a_{1}\zeta /\nu $ is smaller than $a_{2}$ at least by
the factor $1/y$. Assuming inequality $y\gg z$ we can approximate
$a_{2}$ in Eq.\ (\ref{a2a}) by the first expansion term with
respect to $1/y$, $a_{2}\approx -\frac{(6z-1)}{ 2k_{H}^{2}y}$.
Therefore the lower stability boundary is simply given by $z=
1/6$, consistent with the assumed condition $y\gg z$. This is
equivalent to
\begin{equation}
\kappa _{l}(\omega _{E})=\frac{k_{H}}{\sqrt{6}\omega _{E}}.
\label{LowerBoundA}
\end{equation}
This expression for the lower boundary is valid up to $\omega_{E}<
(k_{H}l)^{2/3}/(6\nu_{ab})^{1/3}$.  It is interesting to note that
this boundary does not depend on the dissipation parameters.

Let us consider now the upper boundary $\kappa _{u}(\omega _{E})$.
The term $ a_{1}\zeta /\nu $ in Eq.\ (\ref{StabCondLong1}) can again
be neglected if $\kappa ^{2}\ll \omega _{E}^{2}k_{H}^{2}\nu
_{c}\nu _{ab}$. In this case the upper boundary is again
determined by the equation $a_{2}=0$. At $y\gg 1$ this equation
has asymptotic solution in the form $z=\alpha y\gg 1$ and in the
main order with respect to $1/y$ the constant $\alpha $ is
determined by equation
\[
\mathop{\rm Re} \left[ \frac{i\left( 3\alpha +i\right) }{\left(
\alpha -i\right) ^{3}}\right] =0,
\]
which gives $\alpha =\sqrt{3/5}$. This corresponds to the stability
boundary
\begin{equation}
\kappa _{u}(\omega _{E})=k_{H}\sqrt{\sqrt{\frac{3}{5}}\frac{\nu
_{ab}}{ \omega _{E}}}  \label{UpperBoundA}
\end{equation}
Substituting this expression into the inequality $\kappa ^{2}\ll \omega
_{E}^{2}k_{H}^{2}\nu _{c}\nu _{ab}$ (condition to neglect the term $
a_{1}\zeta /\nu $ in Eq.\ (\ref{StabCondLong1})), we see that
it is equivalent to the condition $\omega _{E}\gg \nu _{c}^{-1/3}$,
which gives the lower limit of applicability of Eq.\
(\ref{UpperBoundA}).

\bigskip
\appendix {\bf Appendix C: Boundary with an insulator}
\medskip

Consider a semiinfinite stack of junctions with $n=1,2,\ldots $
bounded by an insulator at semispace $z<0$ (e.g., by free space).
We assume that the transport current is fed to the stack through
the first layer. In this case the oscillating electric and
magnetic fields induced by the moving Josephson lattice should
match with the electromagnetic oscillations in the insulator.

The oscillating electric and magnetic fields decay into the
insulator as
\begin{eqnarray}
{\bf E}({\bf r},t)={\bf E}_{0}\exp \left( ik_{H}x+qz+i\omega
_{E}t\right),\nonumber \\
{\bf H}({\bf r},t)={\bf H}_{0}\exp \left(ik_{H}x+qz+i\omega _{E}t\right),
\nonumber
\end{eqnarray}
where ${\bf E}_{0}=(E_{x0},0,E_{z0})$, ${\bf H}_{0}=(0,H_{0},0)$,
$q=\sqrt{ k_{H}^{2}-\varepsilon \omega _{E}^{2}/c^{2}}$, and
$\varepsilon $ is the dielectric constant of the insulator. 
Maxwell's equation connects $E_{x0}$ and $H_{0}$ as
\[
-qH_{0}=\frac{i\omega _{E}\varepsilon }{c}E_{x0}.
\]
Continuity of the parallel component of the electric field gives
the relation
\[
E_{x0}=\frac{\Phi _{0}i\omega }{2\pi c}p_{1},
\]
and Eq.\ (\ref{Maxwx}) gives
\[
\left( \frac{2\sigma _{ab}\Phi _{0}}{c^{2}}i\omega _{E}+\frac{\Phi
_{0}}{ 2\pi \lambda _{ab}^{2}}\right)
p_{1}=-\frac{H_{1}-H_{0}}{s}.
\]
From the last three equations we obtain the boundary condition
\begin{equation}
s\sqrt{k_{H}^{2}-\varepsilon \omega
_{E}^{2}/c^{2}}H_{0}=\frac{\varepsilon (\lambda _{ab}\omega
_{E}/c)^{2}}{1+4\pi \lambda ^{2}\sigma _{ab}i\omega
_{E}/c^{2}}\left( H_{1}-H_{0}\right).  \label{BounConEx}
\end{equation}
In general case one has to use the complex dielectric constant 
$\varepsilon$ of the medium at the frequency $\omega_{E}$ and wave 
vector $k_{H}$. In reduced variables this equation can be rewritten as
\begin{equation}
\sqrt{k_{H}^{2}-\frac{\varepsilon }{\varepsilon _{c}l^{2}}\omega
_{E}^{2}}h_{0}=\frac{\varepsilon }{\varepsilon _{c}\gamma
}\frac{\omega _{E}^{2}}{1+i\nu _{ab}\omega _{E}}\left(
h_{1}-h_{0}\right) \label{BounConRed}
\end{equation}
From this relation we obtain the amplitude of reflected wave
\begin{equation}
{\cal B}=-\frac{\sqrt{k_{H}^{2}-\frac{\varepsilon }{\varepsilon
_{c}l^{2}}\omega _{E}^{2}}+\frac{\varepsilon }{\varepsilon
_{c}\gamma }\frac{\omega _{E}^{2}}{1+i\nu _{ab}\omega _{E}}\left(
1-\exp (-iq_{+})\right) }{\sqrt{k_{H}^{2}-\frac{\varepsilon
}{\varepsilon _{c}l^{2}}\omega _{E}^{2}}-\frac{\varepsilon
}{\varepsilon _{c}\gamma }\frac{\omega _{E}^{2}}{1+i\nu
_{ab}\omega _{E}}\left( \exp (iq_{+})-1\right) }
\label{RefGeneral}
\end{equation}
In the case $\varepsilon \sim 1$, $k_{H}\sim 1/\gamma s$, and
$\omega _{E}\sim \omega _{p}$ we obtain the estimate $h_{0}\sim
h_{1}/\gamma \varepsilon _{c}\ll h_{1}$. Therefore the condition
(\ref{BounConRed}) with high accuracy may be replaced by the much
simpler condition $h_{0}=0$, which corresponds to the ideal
reflection case, ${\cal B}=-1$,
\medskip

\bigskip
\appendix {\bf Appendix D: Reflection amplitude for boundary
between moving and static lattice} \medskip

\begin{figure}
\epsfxsize=3.3in \epsffile{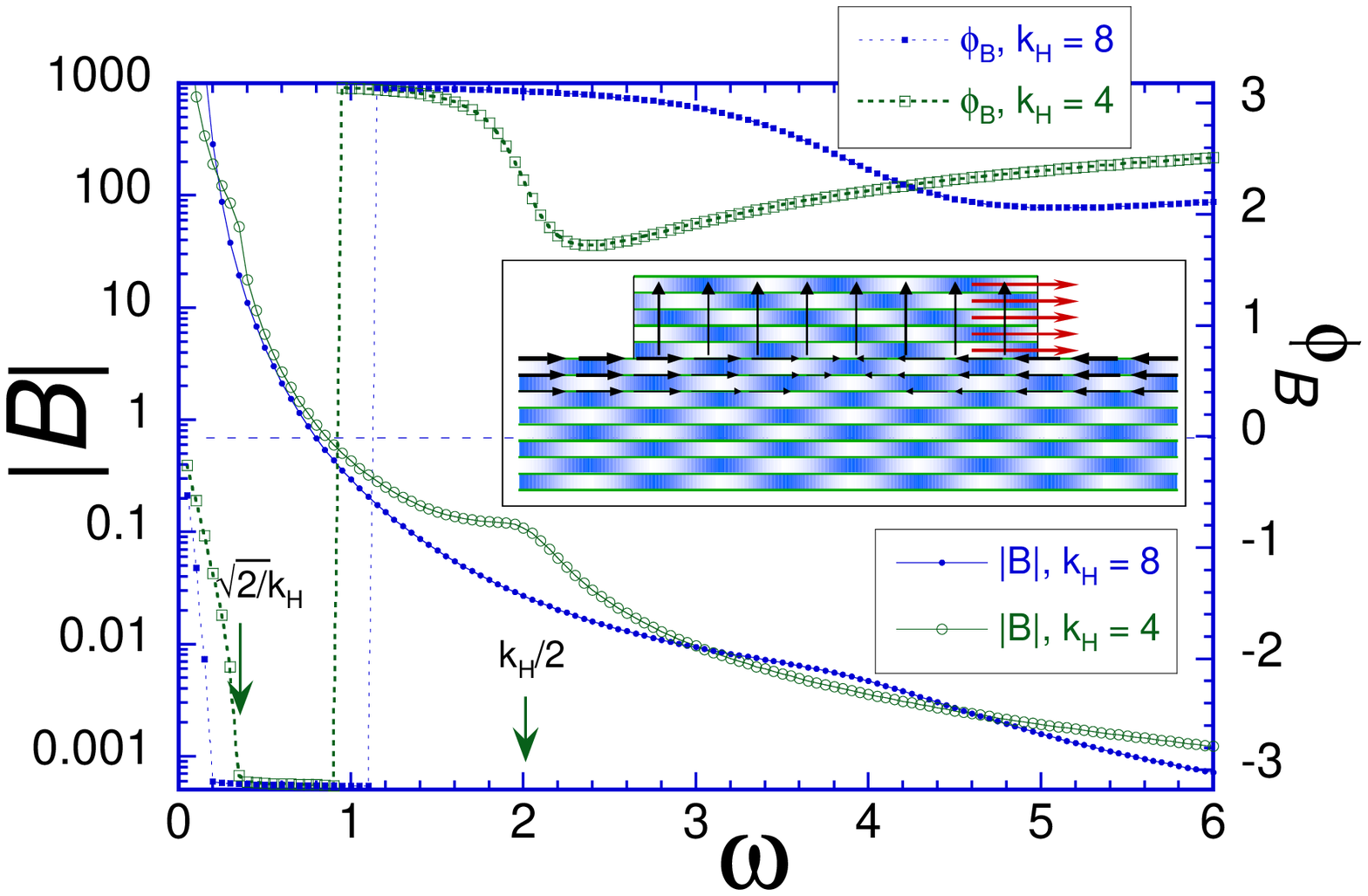} \caption{The frequency
dependence of the absolute value ($|{\cal B}|$) and phase ($\phi_{\cal
B}$) of the amplitude of reflected wave for the boundary between moving
and static lattices (Eq.\ (\protect\ref{AmpStatLat})).  The ploted
curves are computed using parameters $\nu_{c}=0.002$, $\nu_{ab}=0.1$ for
two values of $k_{H}$, $4$ and $8$.  The typical frequencies of the
oscillation spectrum of the static lattice ( $\omega = \sqrt{2}/k_{H}$
and $k_{H}/2$) are marked for $k_{H}=4$.  Inset sketches a small mesa on
the top of bulk crystal.  Concentration of the c-axis transport current
inside the mesa forces the lattice move only in this region.}
\label{Fig-ReflSt}
\end{figure}

Small mesas are typically used in experiment to enhance 
transport current density.  In this case the transport current
flowing in the bulk part of the sample is not sufficient to drive
the Josephson lattice there and the moving Josephson lattice in
mesa neighbors with the static lattice in the bulk (see inset in
Fig.\ \ref{Fig-ReflSt}).  In this appendix we consider
electromagnetic properties of such boundary.  We assume that the
Josephson lattice is static at $n\leq 0$ and moves with constant
velocity at $n>0$. According to the general recipe outlined in
Sec.~\ref{Sec-GenBoun} we have to match electromagnetic
oscillations in the neighboring medium neglecting Josephson
currents in the region of the moving lattice and to find the amplitude of
the reflected wave.  We consider reflection of the wave $\theta
_{n}\propto \exp \left( i\left( k_{H}u+\omega \tau
-iq_{+}n\right) \right)$ coming from above with $q_{+}$ given
by Eq.\ (\ref{qplus}).  This incident wave excites phase and
field oscillations of the static lattice.  Such oscillations have
been theoretically studied in Ref.\ \onlinecite{BulPRB96}. At
high in-plane field these oscillations are described by
approximate equations
\begin{mathletters}
\begin{eqnarray}
\left( \omega ^{2}-i\nu _{c}\omega \right) \theta _{n}-\left( 2{\cal
C}\sin ^{2}\left( k_{H}u+\pi n\right) +\cos \left( k_{H}u+\pi n\right)
\right) \theta _{n}+\frac{\partial h_{n}}{\partial u} &=&0,
\label{OscStata} \\
\left( \nabla _{n}^{2}-\frac{1}{l^{2}}\right) h_{n}+\frac{\partial
\theta _{n}}{\partial u}+i\nu _{ab}\omega \left( \frac{\partial \theta
_{n}}{ \partial u}-\frac{h_{n}}{l^{2}}\right) &=&0, \label{OscStatb}
\end{eqnarray}
\end{mathletters}
where ${\cal C}\equiv \left\langle \cos \theta
_{n}^{(0)}(u)\right\rangle \approx \frac{4+1/l^{2}}{2k_{H}^{2}}$ and
$\theta _{n}^{(0)}(u)\approx k_{H}u+\pi
n-\frac{4+1/l^{2}}{k_{H}^{2}}\sin \left( k_{H}u+\pi n\right) $ is the
static phase difference.  The cosine term in the first equation couples
oscillations with the in-plane wave vector $k_{H}$ to the homogeneous
mode (at $k_{H}\gg 1$ coupling to the higher harmonics can be
neglected).  Eqs.\ (\ref{OscStata}) and (\ref{OscStatb}) have two types
of solutions for the oscillations with the wave vector $k_{H}$, with
$\theta _{n}\propto \cos \left( k_{H}u\right)$ and $\theta _{n}\propto
\sin \left(k_{H}u\right) $, which we refer to as the even and odd modes. 
The incident wave excites both modes but only the even mode is coupled
to the homogeneous oscillations of the lattice.  For this mode we look
for solution in the form $\theta _{n}(u)=\left( \tilde{\theta}\cos
\left( k_{H}u\right) +(-1)^{n}\bar{ \theta}\right) \exp (-i\chi _{e}n)$,
$h_{n}(u)=\tilde{h}\sin \left( k_{H}u\right) \exp (-i\chi _{e}n)$ and
derive equations
\begin{mathletters}
\begin{eqnarray}
\left( \omega ^{2}-i\nu _{c}\omega \right)
\tilde{\theta}+k_{H}\tilde{h} &=&
\bar{\theta}, \\
\left( \omega ^{2}-{\cal C}-i\nu _{c}\omega \right) \bar{\theta}
&=&\frac{
\tilde{\theta}}{2}, \\
\left( 2(1-\cos \chi _{e})+\frac{1+i\nu _{ab}\omega
}{l^{2}}\right) \tilde{h} +k_{H}\left( 1+i\nu _{ab}\omega \right)
\tilde{\theta} &=&0,
\end{eqnarray}
\end{mathletters}
which determine the eigenvalue $\chi _{e}$ for the given frequency
$\omega $ and in-plane wave vector $k_{H}$ as
\begin{equation}
2(1-\cos \chi _{e})=\frac{\left( \omega ^{2}-i\nu _{c}\omega
-{\cal C} \right) \left( 1+i\omega \nu _{ab}\right)
k_{H}^{2}}{\left( \omega ^{2}-i\nu _{c}\omega \right) \left(
\omega ^{2}-i\nu _{c}\omega -{\cal C}\right) -
\frac{1}{2}}-\frac{1+i\omega \nu _{ab}}{l^{2}}.  \label{EvenChi}
\end{equation}
Being inverted, this equation gives the frequencies of plasma modes at
fixed c-axis wave vector $\chi$ and their decay.  Note that at $\omega
=0$ we obtain the solution corresponding to a homogeneous shift of the
lattice, $\chi =\pi $.  For weak dissipation this equation has two other
important solutions: (i) $\omega \approx \sqrt{{\cal C}}\approx
\sqrt{2}/k_{H}$ and $\chi _{e}$ near zero (the endpoint of the
acoustical branch of the oscillation spectrum) (ii) $\omega \approx
k_{H}/2$ and $\chi _{e}\approx \pi $ (the frequency of the homogeneous plasma mode).

For the odd mode we look for solution in the form
$\tilde{\theta}_{n}=\tilde{ \theta}\sin \left( k_{H}u\right) \exp
(-i\chi _{o}n);\;\tilde{h}_{n}=\tilde{h }\cos \left( k_{H}u\right)
\exp (-i\chi _{o}n)$ and derive
\begin{eqnarray*}
\left( \omega _{E}^{2}-i\nu _{c}\omega -{\cal C}\right)
\tilde{\theta}
_{n}-k_{H}\tilde{h} &=&0 \\
-\left( 2(1-\cos \chi _{o})+\frac{1+i\omega \nu
_{ab}}{l^{2}}\right) \tilde{h }+\left( 1+i\omega \nu _{ab}\right)
k_{H}\tilde{\theta} &=&0
\end{eqnarray*}
which gives
\begin{equation}
2(1-\cos \chi _{o})=\left( 1+i\omega \nu _{ab}\right) \left(
\frac{k_{H}^{2} }{\omega ^{2}-i\nu _{c}\omega -{\cal
C}}-\frac{1}{l^{2}}\right) \label{OddChi}
\end{equation}

Because of $\cos (k_{H}u)$ modulation in the static lattice region
the reflected wave contains a backscattered component, with the wave
vector $-k_{H} $, so that the full solution for the oscillating
magnetic field has the form
\begin{eqnarray*}
h &=&\exp (ik_{H}u-iq_{+}n)+{\cal B}\exp (ik_{H}u+iq_{+}n)+{\cal
B}_{b}\exp
(-ik_{H}u+iq_{+}n)\text{, at }n>0 \\
&=&{\cal A}_{e}\cos k_{H}u\exp (-i\chi _{e}n)+i{\cal A}_{o}\sin
k_{H}u\exp (-i\chi _{o}n)\text{, at }n\leq 0
\end{eqnarray*}
Introducing the reflection amplitudes in the even and odd channels,
${\cal B}_{e}= {\cal B}+{\cal B}_{b}$ and$\;{\cal B}_{o}={\cal
B}-{\cal B}_{b}$, we obtain by matching the solutions at $n=0,1$
the following equations
\begin{eqnarray*}
1+{\cal B}_{e} &=&{\cal A}_{e} \\
\exp (-iq_{+})+{\cal B}_{e}\exp (iq_{+}) &=&{\cal A}_{e}\exp
(-i\chi _{e})
\end{eqnarray*}
and similar equations for ${\cal B}_{o}$ and ${\cal A}_{o}$. These
equations can be easily solved giving
\begin{eqnarray*}
{\cal B}_{e} &=&-\frac{\exp (-i\chi _{e})-\exp (-iq_{+})}{\exp
(-i\chi
_{e})-\exp (iq_{+})} \\
{\cal A}_{e} &=&-\frac{2i\sin (q_{+})}{\exp (-i\chi _{e})-\exp
(iq_{+})}
\end{eqnarray*}
and similar formulas for ${\cal B}_{o}$ and ${\cal A}_{o}$.  These
formulas resemble the well-known Fresnel formulas of classical
electrodynamics.  The amplitude of the reflected wave ${\cal B}={\cal
B}(k_{H},\omega )$, which determines the surface contribution to the
response function in Eq.\ (\ref{RespSurfGen}), is given by
\begin{equation}
{\cal B}=-\frac{1}{2}\left( \frac{\exp (-i\chi _{e})-\exp
(-iq_{+})}{\exp (-i\chi _{e})-\exp (iq_{+})}+\frac{\exp (-i\chi
_{o})-\exp (-iq_{+})}{\exp (-i\chi _{o})-\exp (iq_{+})}\right)
\label{AmpStatLat}
\end{equation}
Where, again, $q_{+}$, $\chi _{e}$, and $\chi _{o}$ are given by
Eqs. (\ref{qplus}), ( \ref{EvenChi}), and (\ref{OddChi}).  Plot of
the frequency dependence of the amplitude and phase of ${\cal
B}(k_{H},\omega )$, ${\cal B}(k_{H},\omega )=| {\cal B}|\exp
(i\phi _{{\cal B}})$, is shown in Fig.\ \ref{Fig-ReflSt}.
Typically $|{\cal B}|$ is large in the small frequency range
$0<\omega ^{2}\lesssim {\cal C}$, where the incident decaying wave
excites the propagating acoustic waves in the static lattice.  At
$\omega ^{2}>{\cal C}$ $|{\cal B}|$ rapidly goes to zero, because
both media have identical spectrums of electromagnetic waves at
high frequencies. Both the amplitude and phase of ${\cal
B}(k_{H},\omega )$ have anomalies at the typical frequencies
$\omega = \sqrt{2}/k_{H}$ and $\omega = k_{H}/2$ of the
oscillation spectrum of the static lattice. These anomalies
become more pronounced at lower dissipation and smaller field.

\noindent {\bf Table 1.} Meanings, definitions and practical
formulas for the reduced parameters used throughout the paper. In
practical formulas $f_{p}=\omega_{p}/2\pi$ means plasma frequency,
$\rho_{c}$ and $\rho_{ab}$ are the components of the quasiparticle
resistivity\newline
\begin{tabular}[t]{|c|c|c|c|}
\hline Notation & Meaning & Definition(CGS) & Practical formula
(BSCCO) \\ \hline $\omega _{E}$ & {\ reduced Josephson frequency}
& {\Large {$\frac{2\pi csE_{z}}{\Phi _{0}\omega _{p}}$}} & {\Large
{$\frac{(sE_{z})[{\rm mV}]}{ 2\cdot 10^{-3}f_{p}[{\rm GHz}]}$}} \\
\hline $k_{H}$ & {\ wave vector of Josephson lattice} & {\Large
{$\frac{2\pi H\gamma s^{2}}{\Phi _{0}}$}} &  \\ \hline $\nu _{c}$
& {\ c-axis dissipation parameter} & {\Large {$\frac{4\pi \sigma
_{c}}{\varepsilon _{c}\omega _{p}}$}} & {\Large {$\frac{1.8\cdot
10^{3}}{
\varepsilon _{c}\rho _{c}[\Omega \cdot {\rm cm}]f_{p}[{\rm GHz}]}$}} \\
\hline $\nu _{ab}$ & {\ in-plane dissipation parameter} & {\Large
{$\frac{4\pi \sigma _{ab}\lambda _{ab}^{2}\omega _{p}}{c^{2}}$}} &
{\Large {$\frac{ 0.79(\lambda _{ab}[\mu m])^{2}f_{p}[{\rm
GHz}]}{\rho _{ab}[\mu \Omega \cdot {\rm cm}]}$}} \\ \hline $l$ &
{\ reduced London penetration depth} & $\lambda _{ab}/s$ &  \\
\hline $h$ & {\ reduced local magnetic field} & {\Large
{$\frac{2\pi \gamma \lambda _{ab}^{2}H}{\Phi _{0}}$}} &  \\ \hline
\end{tabular}

\begin{figure}
\epsfxsize=3.3in \epsffile{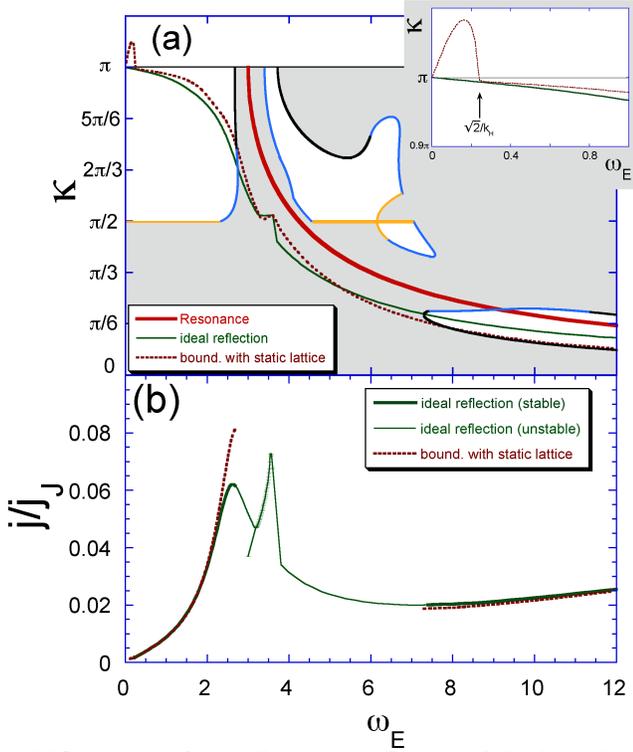} \caption{This figure
illustrates influence of the boundary conditions on the structure
evolution (a) and the current-voltage dependence (b).  We compare the
ideally reflecting boundary (${\cal B}=-1$) and the boundary with
the static lattice (${\cal B}(k,\omega )$ is calculated in Appendix D). 
The comparison is made for $\nu_{c}=0.002$, $\nu_{ab}=0.1$, and $k_{H}=6$. 
Both cases show an overall similar behavior.  In the second case there is a
region at small Josephson frequencies, $0<\omega <\sqrt{2}/k_{H}$, of
the anomalous structure evolution, where the lattice wave vector exceeds
$\pi$ (this region is blown up in the inset).  This region corresponds
to the frequency range of the acoustic branch in the oscillation
spectrum of the static Josephson lattice.  The dependence
$\kappa(\omega_{E})$ has a kink at the endpoint of the spectrum.  However
this point is almost invisible in the I-V dependence.  The second case
is also characterized by the stronger current enhancement near the
instability point (we show only stable I-V branches for this case).  }
\label{Fig-Compar}
\end{figure}

\begin{references}
\bibitem{McCumber} Dissipation in Josephson junctions is
frequently characterized by McCumber parameter $\beta$. It is
connected with $\nu_c$ as $\beta=\nu_c^{-2}$.
\end{references}
\end{document}